\documentclass{jfmplain}

\usepackage{graphicx}
\usepackage{natbib}
\usepackage{graphicx}
\usepackage{xargs}                      
\usepackage[usenames,dvipsnames,svgnames,table]{xcolor}

\usepackage{amsmath}
\usepackage{commath}

\ifCUPmtlplainloaded \elsey
  \checkfont{eurm10}
  \iffontfound
    \IfFileExists{upmath.sty}
      {\typeout{^^JFound AMS Euler Roman fonts on the system,
                   using the 'upmath' package.^^J}%
       \usepackage{upmath}}
      {\typeout{^^JFound AMS Euler Roman fonts on the system, but you
                   dont seem to have the}%
       \typeout{'upmath' package installed. JFM.cls can take advantage
                 of these fonts,^^Jif you use 'upmath' package.^^J}%
      }
  \else
  \fi
\fi

\ifCUPmtlplainloaded \else
  \checkfont{msam10}
  \iffontfound
    \IfFileExists{amssymb.sty}
      {\typeout{^^JFound AMS Symbol fonts on the system, using the
                'amssymb' package.^^J}%
       \usepackage{amssymb}%
         
         \let\geq=\geqslant
      }{}
  \fi
\fi

\ifCUPmtlplainloaded \else
  \IfFileExists{amsbsy.sty}
    {\typeout{^^JFound the 'amsbsy' package on the system, using it.^^J}%
     \usepackage{amsbsy}}
    {}
\fi

\newsavebox{\astrutbox}
\sbox{\astrutbox}{\rule[-5pt]{0pt}{20pt}}

\newcommand{\vx}{\mathbf x}

\newcommand{\vu}{\mathbf u}

\usepackage[colorinlistoftodos,prependcaption,textsize=small]{todonotes}
\newcommandx{\mat}[2][1=]{\todo[inline,linecolor=red,backgroundcolor=red!25,bordercolor=red,#1]{TODO: #2}}
\newcommandx{\marco}[2][1=]{\todo[inline,linecolor=blue,backgroundcolor=blue!25,bordercolor=blue,#1]{TODO: #2}}
\newcommandx{\juan}[2][1=]{\todo[inline,linecolor=OliveGreen,backgroundcolor=OliveGreen!25,bordercolor=OliveGreen,#1]{TODO: #2}}

\title[Mechanisms of Dispersion in a Porous Medium]{Mechanisms of Dispersion in a Porous Medium}
\author[M. Dentz, M. Icardi and J. J. Hidalgo]%
{M. Dentz$^1$%
  \thanks{Email address for correspondence: marco.dentz@csic.es},\ns
M. Icardi$^2$ and J. J. Hidalgo$^1$}

\affiliation{$^1$Spanish National Research Council, IDAEA-CSIC,
c/ Jordi Girona 18, 08034 Barcelona, Spain\\[\affilskip]
$^2$Warwick Mathematics Institute, University of Warwick, CV4 7AL Coventry, UK}

\begin{document}

\maketitle

\begin{abstract}
This paper studies the mechanisms of dispersion in the laminar flow
through the pore space of a $3$--dimensional porous medium. We focus
on pre-asymptotic transport prior to the asymptotic hydrodynamic
dispersion regime, in which solute motion may be described by the
average flow velocity and a hydrodynamic dispersion coefficient. High
performance numerical flow and transport simulations of solute
breakthrough at the outlet of a sand-like porous medium evidence
marked deviations from the hydrodynamic dispersion paradigm and
identify two distinct regimes. The first regime is
characterized by a broad distribution of advective residence times in
single pores. The second regime is characterized by diffusive mass
transfer into low-velocity regions in the wake of solid grains. These
mechanisms are quantified systematically in the framework of a
time-domain random walk for the motion of marked elements (particles)
of the transported material quantity. Particle transitions occur over
the length scale imprinted in the pore structure at random times given
by heterogeneous advection and diffusion. Under globally
advection-dominated conditions, this means P\'eclet numbers larger than
$1$, particles sample the intrapore velocities by diffusion, and the
interpore velocities through advection. Thus, for a single transition,
particle velocities are approximated by the mean pore velocity. 
In order to quantify this advection mechanism, we develop a model for the
statistics of the Eulerian velocity magnitude based on Poiseuille's law
for flow through a single pore, and for the distribution of mean pore
velocities, both of which are linked to the distribution of pore
diameters. Diffusion across streamlines through immobile zones in the
wake of solid grains gives rise to exponentially distributed residence
times that decay on the diffusion time over the pore length. The
trapping rate is determined by the inverse diffusion time. This
trapping mechanism is represented by a compound Poisson process conditioned on the
advective residence time in the proposed time-domain random walk
approach. The model is parameterized with the characteristics of the
porous medium under consideration and captures both
pre-asymptotic regimes. Macroscale transport is described by an
integro-differential equation for solute concentration, whose memory
kernels are given in terms of the distribution of mean pore velocities
and trapping times. This approach quantifies the physical non-equilibrium caused by
a broad distribution of mass transfer time scales, both advective and
diffusive, on the representative elementary volume (REV). Thus, while
the REV indicates the scale at which medium properties like porosity
can be uniquely defined, this does not imply that transport
can be characterized by hydrodynamic dispersion.

\end{abstract}


\section{Introduction\label{sec:intro}}
Transport of a dissolved substance or heat in the laminar flow through the
void and pore space of a porous medium is due to molecular diffusion
and heterogeneous advection induced by the complex pore
structure. These mechanisms give rise to the asymptotic phenomenon of
mechanical or hydrodynamic dispersion~\cite[][]{Bear:1972}. In order to
illustrate this phenomenon, we take the view point of marked elements of
the transported material quantity, or idealized solute
particles, whose density is equivalent to the solute distribution, and
whose motion is governed by advection and a stochastic velocity that
represents diffusion~\cite[][]{Risken:1996,Gardiner:2009}. Dispersion
quantifies the impact of velocity fluctuations on particle transport, similar to the concept of
Brownian motion. In this context velocity fluctuations
occur on a fixed characteristic time scale. Thus, particles have access
to the full fluctuation spectrum at each moment, or in other words,
they are statistically equal. The system is in local
physical equilibrium. For transport in a porous medium, this
is in general different. Velocity fluctuations occur on characteristic
length scales imprinted in the porous medium structure. Particles'
residence times in regions of small velocities are larger than in
regions of high velocities~\cite[][]{saffman1959theory}. As a
consequence, statistical equivalence between particles and thus local
physical equilibrium is achieved only asymptotically for times much
larger than the largest residence time. 

The concept of residence times has been used in the pioneering works of
\cite{josselin1958} and \cite{saffman1959theory}.    
These authors studied hydrodynamic dispersion and its mechanisms in the
light of pore-scale advection and diffusion in order to determine the
 longitudinal (in mean flow direction) and transverse dispersion coefficients.
Their approaches are based on modeling particle transport
as what is now known as a time-domain or continuous time random
walk~\cite[][]{SL73.1,Cvetkovic1991,noetinger2016}, in which particles
perform transitions over a characteristic pore-length with transition times that
depend on both advection and diffusion. For example, if the advection time over a
pore is larger than the diffusion time, the residence time is given by
the characteristic diffusion time~\cite[][]{saffman1959theory}. \cite{Bijeljic2006}
consider pore network models characterized by broad distributions of
advective residence times, which are cut-off at the
diffusion time, in order to explain the
dependence of longitudinal hydrodynamics dispersion coefficients on
the P\'eclet number~\cite[][]{Pfannkuch1963}. The P\'eclet number
compares the diffusion to the advection time over a characteristic
distance. Recent experimental and numerical works have focused on the
quantification of the P\'eclet dependence of longitudinal and
transverse hydrodynamic dispersion~\cite[][]{scheven2013pore,
  icardi2014pore}.  

For purely advective pore-scale transport, observed anomalous or
non-Fickian transport patterns have been modeled using the continuous
time random walk approach~\cite[][]{bijeljic2011signature,
  bijeljic2013, DeAnna2013, lester2014, Kang2014, Holzner2015} based on broad
distributions of advective residence times. Recent numerical and
experimental works, which studied purely
advective particle transport through porous media, have uncovered
intermittent patterns in particle
velocities and accelerations as a consequence of broad distributions
of flow velocities and their spatial organization~\cite[][]{DeAnna2013,Kang2014,Holzner2015}. 
Other works~\cite[][]{siena2014relationship,gjetvaj2015,Jin2016,Matyka2016}
have studied the distribution of flow velocities and their relation to the 
pore size in synthetic porous media. These works show that, in the
absence of molecular diffusion, transport may be persistently
anomalous, characterized by non-linear growth of the longitudinal
centered mean square displacement and tailing of the arrival time distributions of
particles at a control plane. Such behaviors are caused mainly by low
flow velocities as may occur close to the grains and through small
pores. 

As pointed out by \citet{saffman1959theory}, the impact of
diffusion on particle transport and transport velocities is
two-fold. On one hand, it homogenises the flow velocities in single
pores as particles may sample velocities across different
streamlines. This is similar to particle diffusion in the flow through a
circular pipe~\cite[][]{Taylor:1953}, for which the mean particle
velocities asymptotically approach the mean flow velocity. On the
other hand, as mentioned above, diffusion provides a cut-off mechanism
if the advective transition time is much larger than the
characteristic diffusion time scale over a pore. Thus, depending on the
P\'eclet number, the time scales to reach the asymptotic regime of
hydrodynamic dispersion may be very large compared to the 
advection time across a representative elementary
volume (REV)~\cite[][]{Bear:1972} by the average flow velocity. The
REV comprises enough pore-lengths such that macroscopic medium
properties as porosity and permeability, for example, can be uniquely
defined. It is typically assumed that transport on the REV scale can
be represented in terms of the average flow velocity and hydrodynamic
dispersion coefficients~\cite[][]{Bear:1972,Whitaker:book,Hornung:book}, called
advection-dispersion approach in the following. However, depending on
the P\'eclet number, and on the pore-scale mass transfer mechanisms,
for example between mobile and immobile
porosity~\cite[][]{GMDC2008, liu2012applicability, gjetvaj2015},
this may not be the case.   


Current macroscale models of pore-scale transport
include the advection-dispersion equation, which however, may
only be valid at asymptotic times, or continuous time random walks
for the modeling of purely advective non-Fickian
transport. The relation between pore structure and flow velocities,
and the role of advection and diffusion in preasymptotic transport
remain open research questions.  

\begin{figure}
\makebox[\textwidth][c]
{
\includegraphics[width=.7\textwidth]{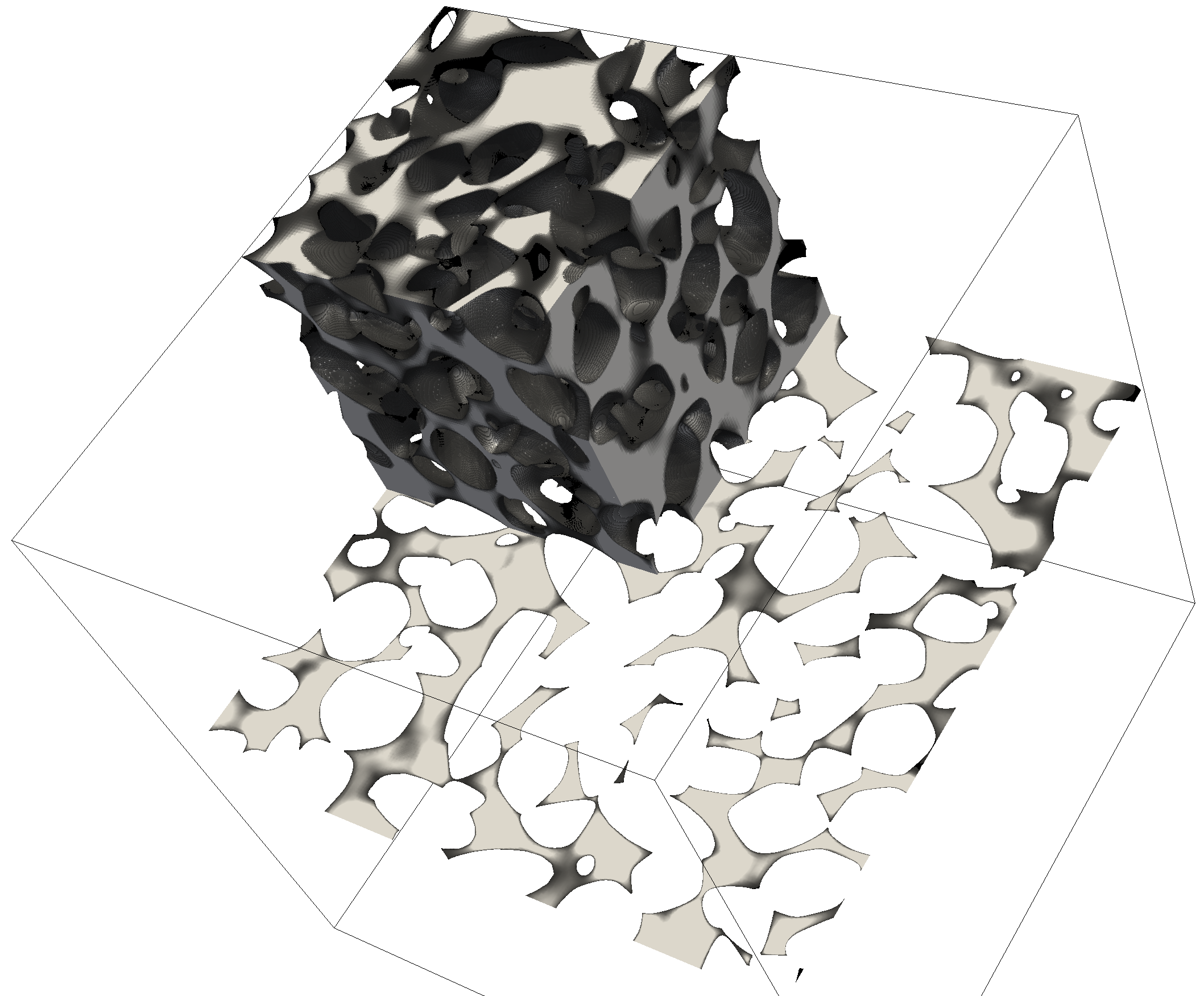}
}
\caption{Three-dimensional (one eighth) sample of the pore space of the
  medium under investigation and a two-dimensional slice of it.}
\label{fig:2d3d}
\end{figure}

We address these questions for advective-diffusive solute transport in
a $3$--dimensional porous medium similar in grain size distribution
and porosity to sands or bead packs~\cite[][]{icardi2014pore}, see also
Figure~\ref{fig:2d3d}. High performance computational fluid
dynamics simulations of flow and transport provide the data of
preasymptotic solute transport, which evidences deviations from the
advection-dispersion behavior and its dependence on the P\'eclet
number. Based on the approach of \citet{saffman1959theory}, which
relates the pore-diameter to the maximum pore velocity through
Poiseuille's law, we propose a link between the pore-size distribution 
and the distribution of the Eulerian velocity magnitudes. We then
formulate particle transport as a time-domain random walk with
particle transitions over a characteristic pore length and temporal
transitions corresponding to the dominant advection and diffusion
mechanisms. We first consider purely advective transitions, which,
however, fail to describe the numerical data. We successively identify
and quantify the interaction of the intra and interpore particle
advection and diffusion mechanisms. These include homogenization of
intrapore velocities due to diffusion across streamlines, diffusion as
a transport mechanism along streamlines~\cite[][]{saffman1959theory}
and diffusion across streamlines into low velocity zones in the wake
of solid grains. These mechanisms are integrated systematically into
an upscaled transport model on the basis of time-domain random walks,
and parameterized in terms of the statistical medium and flow
properties. 

The paper is organized as follows. In Section~\ref{sec:pore-scale} the
pore-scale flow and transport problem is described and the medium
structure and numerical set-up are given. The simulation results for
solute breakthrough curves are discussed in the light of the
advection-dispersion approach. Section~\ref{sec:model} analyses the
pore-scale velocity distribution and proposes a model for the velocity
magnitude and the distribution of mean pore velocities 
in terms of the distribution of pore-diameters. Section~\ref{sec:CTRW}
studies the mechanisms of pore-scale advection and diffusion and their
impact on pre-asymptotic particle transport. These mechanisms are cast
into a time-domain random walk model, which is compared against
the breakthrough curves from the detailed numerical flow and transport
simulations. 


\section{Pore Scale Flow and Transport\label{sec:pore-scale}}

Flow and transport at the pore-scale are governed by 
the following Stokes equation for the flow velocity $\vu(\vx)$ 
and the advection-diffusion equation for the conservative scalar $c(\vx,t)$,
\begin{align}
\label{stokes}
& \mu \nabla^2 \mathbf u(\vx) = \nabla p(\vx)
\\
\label{ade}
&\frac{\partial c(\vx,t)}{\partial t} + \nabla \cdot \mathbf u(\vx,t) c(\vx,t)
  - D \nabla^2 c(\vx,t) = 0,
\end{align}
where $\mu$ is viscosity, and $D$ is the molecular diffusion
coefficient. These equations are solved numerically with the open-source
finite-volume code \textsf{OpenFOAM 4.x}. Details about numerical
discretisation and solvers are discussed in Appendix~\ref{app:of} while
boundary conditions and operating conditions are discussed below.
We employ here a fully Eulerian formulation that, provided an accurate
discretisation and finite diffusion coefficient, is robust and has no
statistical error.

\subsection{Model medium}
We consider the pore-scale sample described in
\citet{icardi2014pore}. As detailed in this study, the synthetic medium was generated
according to the characteristics of standard sand samples. The grain
size distribution was obtained from scanning electron microscopy and
static-light scattering measurements and fitted by a Weibull
probability density function (PDF) for the diameter of equivalent spheres,
\begin{align}
\label{pgrain}
p_d(d) =
  \frac{k}{d_0}\left(\frac{d}{d_0}\right)^{k-1}e^{-(d/d_0)^{k}}. 
\end{align}
The mean grain diameter is $\langle d \rangle = d_0 \Gamma(1+1/k) =
0.277$ mm and the Weibull parameter is $k = 7$. This distribution is
sharply peaked about its mean. The synthetic porous medium was
generated by sampling grain sizes from the Weibull
distribution~\eqref{pgrain} and sedimentation of 
irregularly shaped grains using the software package
BLENDER, which resulted in a porosity of $n=0.35$.
Note that the Weibull distribution~\eqref{pgrain} represents a
parametric PDF to represent the empirical grain size
distribution. The lognormal PDF is another frequently used
parameterization for empirical grain size
distributions~\cite[][]{Friedman}.  
The model medium is cubic of length $L=2$mm. It contains
about $2 \cdot 10^3$ grains. The characteristic pore length
$\ell_0$ is of the order of the mean
grain size $\ell_0 \approx \langle d \rangle = 0.277$ mm. The length
$L$ of the study domain is $L = 7.22
\langle d \rangle$. The volume is of the size of a representative
elementary volume in terms of the definition of volumetric
porosity~\cite[][]{Bear:1972}. This means in terms of this Darcy-scale property,
the medium can be considered macroscopically
homogeneous. Figure~\ref{fig:2d3d} shows a section of the
$2$-dimensional porous medium and a $2$-dimensional slice. 

A uniform pressure boundary conditions is imposed at the
inlet and the outlet, while zero-flux boundary conditions are imposed
on the lateral boundaries. This results in a mean velocity $\langle
\vu(\vx) \rangle \equiv \langle u_1 \rangle \mathbf e_1$
oriented towards the $1$-axis of $\langle u_1 \rangle =
5.73\cdot10^{-6}$ m/s and
identifies the characteristic advection time $\tau_u = L / \langle u_1 \rangle$. The P\'eclet
number, which compares advective and diffusive transport over a pore length $\ell_0$, is defined by 
\begin{align}
\label{Pe}
Pe = \frac{\langle u_1 \rangle \ell_0}{D}. 
\end{align}
%
\begin{figure}
\makebox[\textwidth][c]
{
\includegraphics[width=.45\textwidth]{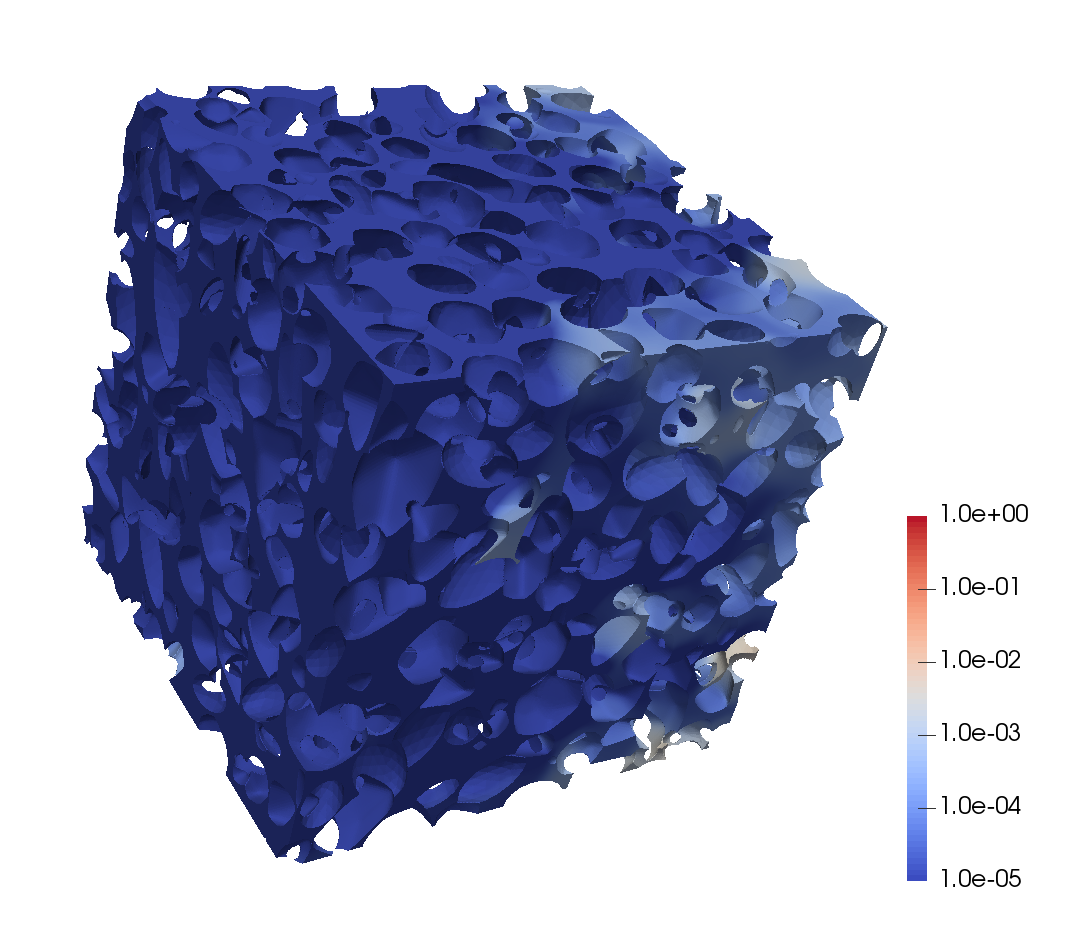}
\includegraphics[width=.45\textwidth]{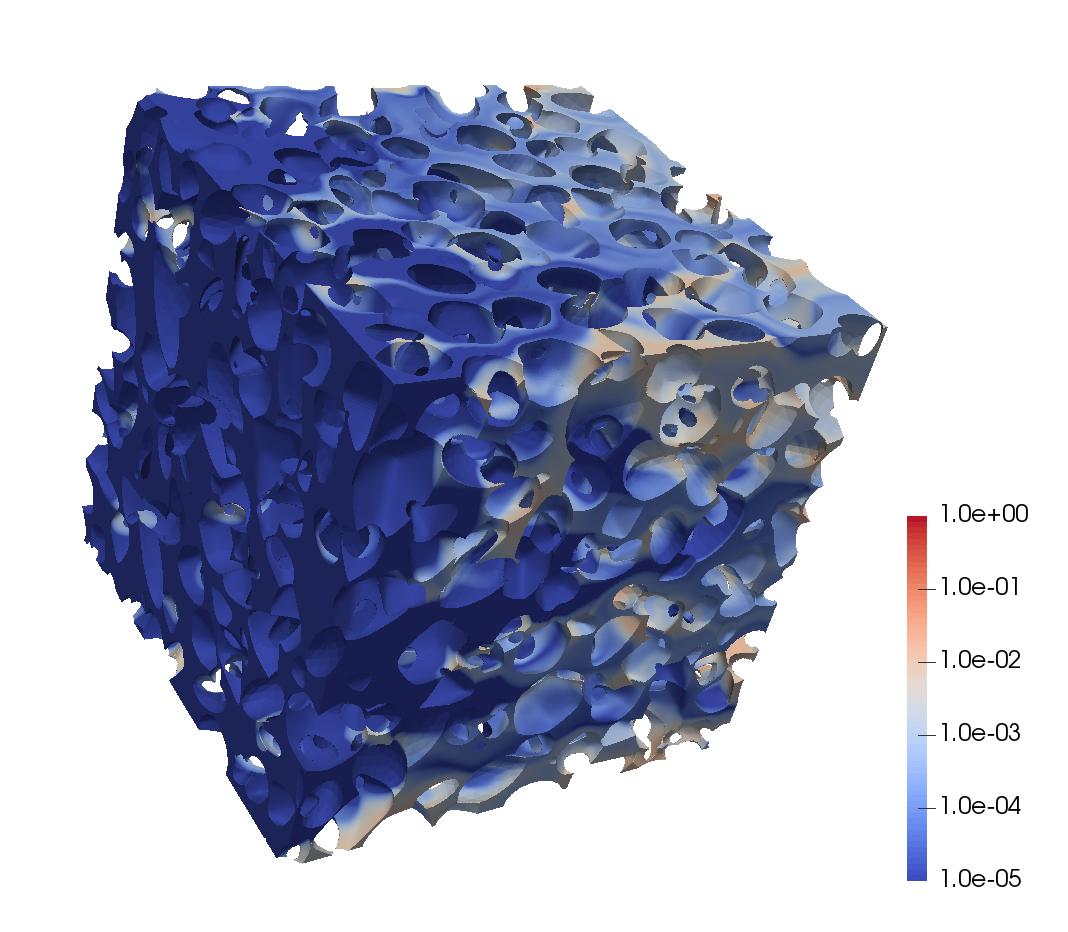}
}
\caption{Three-dimensional snapshot of the concentration
  approximately $t \approx \tau_u/2$ after a pulse injection for
  P\'eclet 30 (left) and 1000  (right). Flow is from left to right.}
\label{fig:snapshots}
\end{figure}

\subsection{Solute transport}

\begin{figure}
a.\includegraphics[width = .45\textwidth]{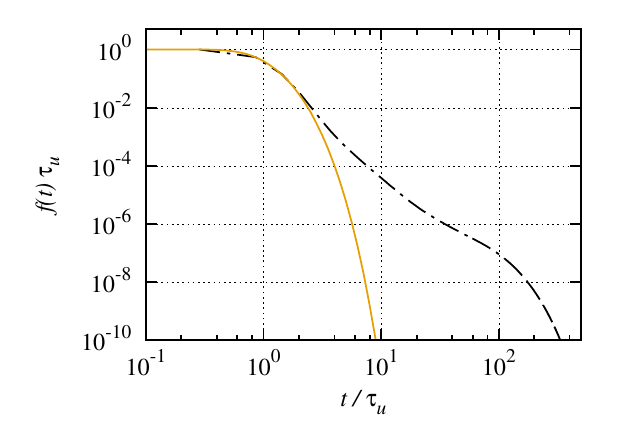}
b.\includegraphics[width = .45\textwidth]{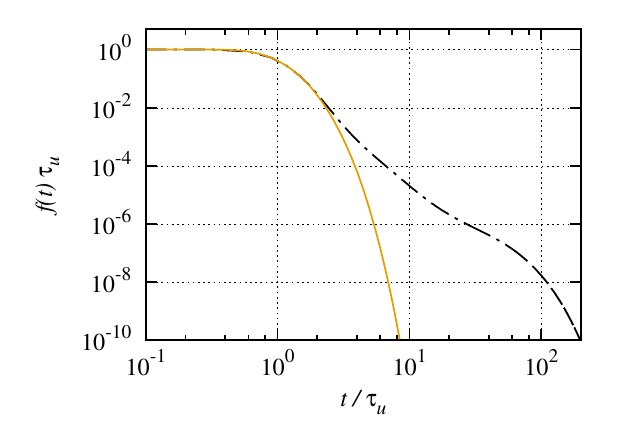}

c.\includegraphics[width = .45\textwidth]{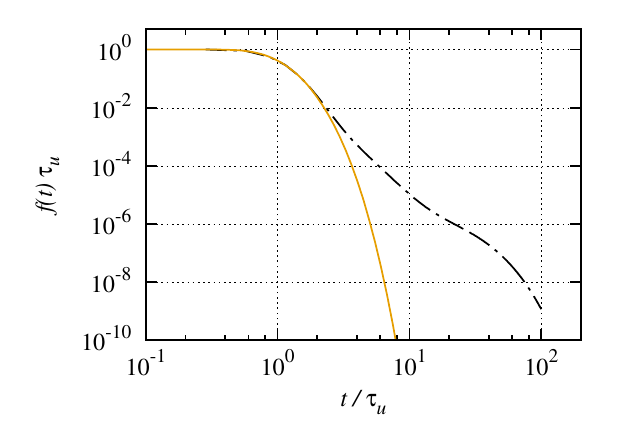}
d.\includegraphics[width = .45\textwidth]{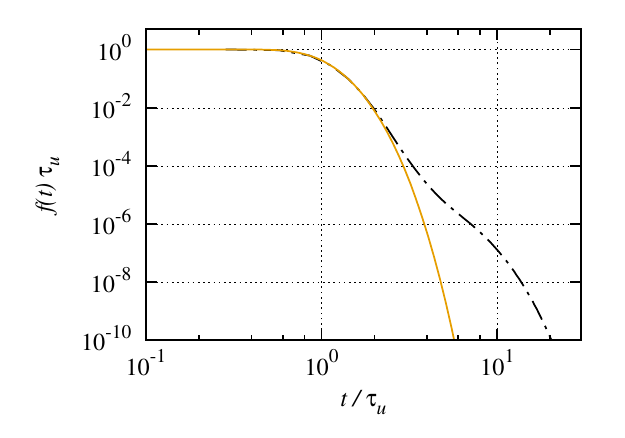}
\caption{Breakthrough curves for $Pe = 10^3$, $5 \cdot 10^2$, $3 \cdot
  10^2$ and $30$. The dash-dotted lines in the following indicate the
  numerical data. The solid lines denote the solution~\eqref{eq:ade_analytic} characterized by the
  dispersion coefficient $\mathcal D$ which is directly fitted from
  the breakthrough data. 
\label{fig:btc}}
\end{figure}
The solute transport equation~\eqref{ade} is solved in a fully
Eulerian way. Different microscopic P\'eclet numbers, defined
by~\eqref{Pe}, are imposed by tuning appropriately the
molecular diffusion coefficient $D$. Note that $Pe$ is the only
dimensionless group for the transport problem under
consideration. Thus, varying the mean flow rate and varying the
diffusion coefficient have the same effect as long as $Pe$ remains
unchanged. No-flux boundary conditions are imposed at all boundaries but the inlet where a constant
concentration is assumed. Figure~\ref{fig:snapshots} shows snapshots
of the concentration distribution in the pore space. We observe
increasing spatial heterogeneity with increasing P\'eclet
number. Specifically, for the high P\'eclet, the invasion front is not
homogeneous with regions of low solute concentration next to regions
of high concentration. This is an expression of local physical
non-equilibrium as discussed below. 

Transport is characterized in terms of the
complementary cumulative breakthrough curve defined as 
\begin{align}
f(t) = 1-\frac{\displaystyle\int dx_2 \int dx_3 c(x_1=L,x_2,x_4,t) u_{1}(x_1=L,x_2,x_3)}
 {\displaystyle \int dx_2 \int dx_3 u_{1}(x_1=L,x_2,x_3)}. 
\end{align}
In the following, we refer to $f(t)$ as the breakthrough
curve. Figure~\ref{fig:btc} shows the solute breakthrough curves for
different P\'eclet number. We observe increased tailing with
increasing $Pe$ in line with the snapshots of the concentration
distribution shown in Figure~\ref{fig:snapshots}. Tailing of the
breakthrough curves is an expression of local non-equilibrium because
it is the result of a broad distribution of residence times in the
sample. 
 
In order to understand and interpret the breakthrough behaviour, we first consider the classical advection-dispersion
approach. \citet{icardi2014pore} quantified the cumulative
complementary breakthrough curves in the light of this approach, which characterises transport in
terms of an advection dispersion equation (ADE) for the evolution of the
macroscale concentration $\overline c(x,t)$~\cite[][]{Whitaker:book, Hornung:book}
\begin{align}
\label{ademac}
\frac{\partial \overline c(x,t)}{\partial t} + \langle u_1\rangle \frac{\partial
  \overline c(x,t)}{\partial x} - \mathcal D \frac{\partial^2 \overline c(x,t)}{\partial x^2}
  = 0, 
\end{align}
where $\mathcal D$ is the hydrodynamic dispersion coefficient. For
simplicity in the following, we denote $x = x_1$. Such a 
macroscale transport equation may be derived using volume averaging
or homogenisation theory~\cite[][]{Whitaker:book, Hornung:book}. It is
a valid transport description if the support scale, the representative
elementary volume~\cite[][]{Bear:1972} is well mixed. Under well-mixed
conditions, the macroscale concentration $\overline c(x,t)$ uniquely
determines the concentration on the support scale at the
position $x$~\cite[][]{mixing:book}, which is not the case here for
large $Pe$. The solution for the
complementary cumulative breakthrough curve $f(t,x)$ at a position $x$
in a semi-infinite medium is then given by~\cite[][]{KreftZuber1978} 
\begin{align}
\label{eq:ade_analytic}
f(t,x) = 1-\frac{1}{2} \left[\text{erfc}\left(\frac{x-t}{\sqrt{4
  \mathcal D t}}\right)+\exp\left(\frac{x}{\mathcal D}\right)\text{erfc}\left(\frac{x+t}{\sqrt{4
  \mathcal D t}}\right)\right]. 
\end{align}
The latter is a good approximation to the solution for a finite domain
for $Pe \gg 1$, which is fulfilled for the scenarios of interest
here. We denote $f(t) = f(t,x = L)$. 
Figure~\ref{fig:btc} compares~\eqref{eq:ade_analytic} to the numerical data for
different $Pe$. The onset and initial decay of the breakthrough
curve are well represented by the mean flow velocity and a fitted 
dispersion coefficient $\mathcal D$. However, it fails to capture the
tailing behaviour at intermediate and long times. This implies that
the system is not at local equilibrium, the support scale is not well-mixed.
In the following we analyse these transport features and their origins in terms of spatial
fluctuations of the pore-scale flow velocity and the effect of molecular
diffusion. The aim is to gain understanding of the pore-scale transport
mechanisms in order to arrive at a macroscale transport approach that
can be quantified in terms of the pore-scale velocity, pore size
distributions and molecular diffusion. We first discuss the statistical properties of the
Eulerian velocity and velocity magnitude in the light of the pore-size
distribution before we analyze the mechanisms of pore-scale transport and
their macroscale quantification in terms of $f(t)$. 
\section{Eulerian Velocity Distributions and Pore Scale Velocity Model\label{sec:model}}
In order to understand the pore-scale dispersion behaviour, we 
study here the Eulerian velocity distribution and its interpretation in
terms of the grain and pore-size distribution. We consider the
distribution of the velocity components in mean flow direction,
$u_1(\vx)$, and perpendicular to it, $u_i(\vx)$ for $i \neq 1$, as
well as the velocity magnitude 
%
$v_e(\vx) = \sqrt{\sum_i u_i(\vx)^2}$.   
%
The Eulerian velocity PDFs are computed as volume weighted histograms
over the entire pore-space $\omega$
\begin{align}
\label{pev}
p_e(v) = \frac{1}{V_\omega} \int\limits_{\omega} d \vx \delta[v - v(\vx)],
\end{align}
where $V_\omega$ is the pore volume; $v(\vx) = u_i(\vx)$ for the PDF of velocity components and
$v(\vx) = v_e(\vx)$ for the PDF of the velocity magnitude. Note that
these PDFs generally depend on the sampling volume. For
the sand-like porous medium under consideration here, the sampling
volume is of the size of an REV, thus the velocity PDF is assumed not
to change significantly when increasing the sampling volume.  

\begin{figure}
\begin{center}
\includegraphics[width=.45\textwidth]{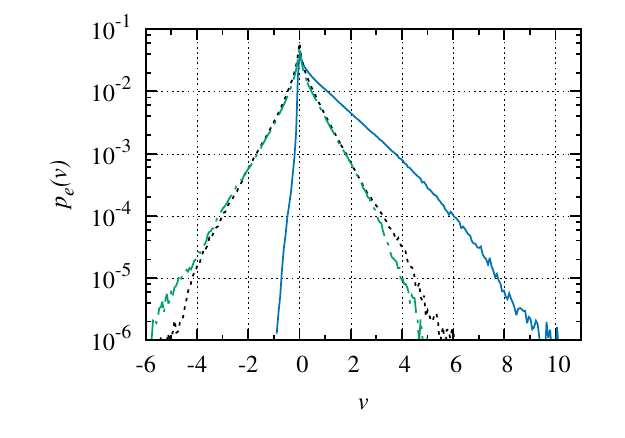}
\includegraphics[width=.45\textwidth]{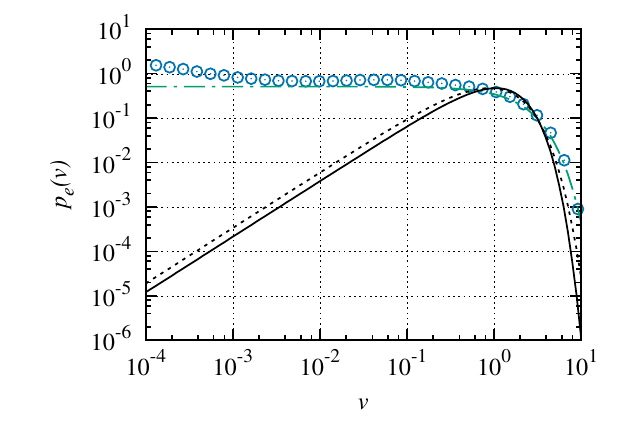}
\end{center}
\caption{(Left) PDF of (solid) $u_1(\vx)$ (dashed) $u_2(\vx)$ and
  (dash-dotted) $u_3(\vx)$.
  (Right) (circles) PDF of the magnitude of Eulerian
  velocities and (dash-dotted) estimation with the velocity
  model~\eqref{pe} for $v_0 = 2.1 \langle u_1 \rangle$ and $\alpha = 2.5$. The solid line
  denotes the PDF~\eqref{pmv} of mean pore velocities, the dashed line
  its approximation by~\eqref{pmgamma}. 
  Velocities are normalized by $\langle u_1 \rangle$. 
\label{fig:vpdf}}
\end{figure}

The PDFs of the
velocity components are shown in Fig.~\ref{fig:vpdf}. The distributions of the
components perpendicular to the mean flow direction are symmetric
around $0$ and have an exponential shape. The PDF of the longitudinal
component is skewed towards positive values with an approximately
stretched exponential shape~\citep{siena2014relationship,Holzner2015}. It is
non-zero for small negative velocities, which indicates the presence
of back-flow. It  is not clear how to relate the distributions of the
velocity components to statistical characteristics of the porous medium. 

In order to relate the medium properties to the Eulerian flow field,
we focus rather on the PDF of velocity magnitudes illustrated in
Figure~\ref{fig:vpdf}. The PDF is flat from $10^{-3}$ to $1$, from
which on it decays exponentially fast. As detailed below, the flat
distribution at small velocities can be understood by sampling a
parabolic velocity profile, the smooth cut-off at larger velocities
may be related to the distribution of pore-sizes. The deviation from
the flat profile at small velocities may be attributed to low
velocity zones in the wake of solid grains as discussed in
Section~\ref{sec:mrmt}.

First, we consider the mean of the Eulerian velocity magnitude. It is related to the mean
velocity through the advective tortuosity
$\chi$ as~\cite[][]{koponen1996tortuous}
\begin{align}
\langle v_e \rangle = \chi \langle u_1\rangle.  
\end{align}
The advective tortuosity measures the ratio of average trajectory length
to linear distance. For the model medium under consideration, the mean
flow velocity is $\langle u_1 \rangle = 5.7 \cdot 10^{-6}$ m/s, and the
mean velocity magnitude $\langle v_e \rangle = 9 \cdot 10^{-6}$
m/s. Thus, the tortuosity is given by $\chi = \langle v_e \rangle /
\langle u_1 \rangle = 1.6$.  

\begin{figure}
\begin{center}
\includegraphics[width=\textwidth]{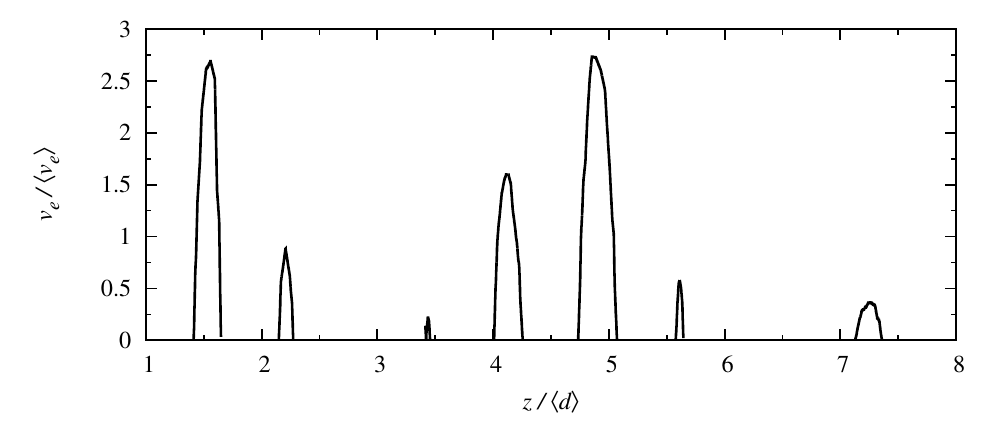}
\caption{Velocity profile across the porous medium perpendicular to the main flow
  direction. }
\label{fig:profiles}
\end{center}
\end{figure}

In the following, we study the PDF of velocity
magnitudes based on the assumption that the velocity profile in a single pore is
parabolic~\citep{saffman1959theory,
  lester2013chaotic,Holzner2015,deannaprf2017}. The validity of this
assumption is underlined qualitatively by the velocity profiles shown
in Fig.~\ref{fig:profiles}, which are obtained along a cut
of the three-dimensional porous medium of Fig.~\ref{fig:snapshots}. 
Note that some authors~\cite[][]{LeBorgneWRR2011} have
studied flow and transport in wavy tubes, this means a serial
arrangement of pores, as a model system for the
upscaling of pore-scale heterogeneity. This view is not appropriate
for more complex porous media, for which the network aspect is
important in the sense that pore intersections exist such that the concept of the
linear pore breaks down. Some aspect of flow in serial and
parallel pore arrangements are discussed
in~\cite[][]{Holzner2015}.  

The PDF~\eqref{pev} of the velocity magnitude can be decomposed into
the contributions from the individual pores as
\begin{align}
p_e(v) = \sum_p \frac{V_p}{V_\omega} \frac{1}{V_p}
  \int\limits_{\omega_p} d\vx  \delta[v -
  v(\vx)],
\end{align}
where $\omega_p$ is the single pore domain and $V_p$ its
volume. The velocity $v(\vx) \equiv v(r)$ depends only on the pore
radius and is given by
\begin{align}
\label{vr}
v(r) =  v_p \left[1 - \left(\frac{2 r}{a_p} \right)^2 \right], 
\end{align}
where $a_p$ is the pore diameter and $r$ is the distance from the pore
axis. The maximum velocity $v_p$ in a pore
is given by $v_p = v_0 (a_p/a_0)^2$ with $v_0$ a characteristic
velocity and $a_0$ a characteristic pore diameter. Under this assumption, we obtain
\begin{align}
p_e(v) = \sum_p \frac{V_p}{V_\omega} \frac{1}{v_p} H(v_p - v),
\end{align}
where $H(v)$ denotes the Heaviside step function. 
Note that the velocity PDF in a single pore is uniform between $0$ and
the maximum velocity $v_p$, which can be seen by sampling
$v(r)$ given by~\eqref{vr} uniformly in a pore
cross-section. 

We furthermore assume that the pore volume is $V_p
= b a_p^2 \ell_p$ with $b$ a shape factor, $\ell_p$ the pore length
and $a_p$ the pore diameter. In the following, we assume that the pore
length $\ell_p \approx \ell_0$ is approximately constant, while the
pore diameter $a_p$ is variable. For the sand-like medium under
consideration, the pore-length is of the order of the grain size. As
the grain size distribution~\eqref{pgrain} is sharply peaked, we
approximate it as constant. The pore diameter in contrast depends on the
distance of the grains in the packing, which is much more
variable. This is indicated in the two-dimensional slice of the
porous medium shown in Figure~\ref{fig:2d3d}.  
We can now write
\begin{align}
p_e(v) = \sum_p \frac{b a_0^2 \ell_0}{V_\omega} \frac{1}{v_0} H[v_0
  (a_p/a_0)^2 - v].
\end{align}
This expression can be written tautologically as 
\begin{align}
\label{pev1}
p_e(v) = \int d a \frac{b a_0^2 \ell_0 N_0}{V_\omega} \frac{1}{v_0} H[v_0
  (a/a_0)^2 - v] \left[\frac{1}{N_0}\sum_p \delta(a-a_p)\right],
\end{align} 
where $N_0$ is the number of pores. We note that the expression in square brackets
on the right side is equal to the PDF of pore diameters $p_a(a)$, and
define the average volume of a single pore as $\langle V_p
\rangle = V_\omega/N_0$. Thus, expression~\eqref{pev1} can be written
as
\begin{align}
\label{pev2}
p_e(v) = \int d a \frac{b a_0^2 \ell_0}{\langle V_p \rangle} \frac{1}{v_0} H[v_0
  (a/a_0)^2 - v] p_a(a),
\end{align}
%

\begin{figure}
\begin{center}
\includegraphics[width = .7\textwidth]{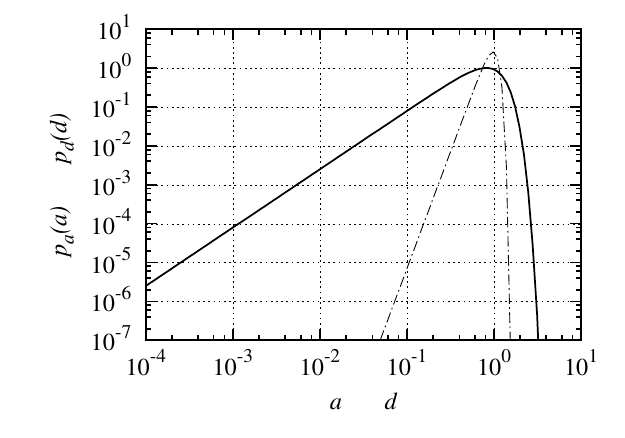}
\end{center}
\caption{The dash-dotted line illustrates the PDF of grain diameters~\eqref{pgrain}
  for $d_0 = 1$, the solid line shows the corresponding PDF of pore
  diameters~\eqref{weibull} for $a_0 = 1$ obtained from the proposed velocity model.\label{fig:aPDF}}
\end{figure}

The grain size distribution is given by a Weibull distribution. We assume
for the pore-size distribution also a Weibull distribution with 
\begin{align}
\label{weibull}
p_a(a) = \frac{\alpha}{a_0} \left(\frac{a}{a_0}\right)^{\alpha-1}
  \exp[-(a/a_0)^\alpha]. 
\end{align}
We note that the average pore volume is $\langle V_p \rangle \propto
a_0^2 \ell_0$. Using these relations in~\eqref{pev2}, we obtain after
integration and normalisation the following estimation for the
Eulerian velocity PDF, 
%
\begin{align}
\label{pe}
p_e(v) = \frac{\exp\left[-(v/v_0)^{\alpha/2}\right]}{v_0 \Gamma(1+2/\alpha)} . 
\end{align}
The mean velocity is given by 
\begin{align}
\langle v_e \rangle = v_0 \frac{\Gamma(4/\alpha)}{\Gamma(2/\alpha)}. 
\end{align}
By fitting the Eulerian velocity data, we estimate an exponent of
$\alpha \approx 2.5$ and $v_0 \approx 2.1 \langle u_1 \rangle$, which has the same average
velocity of $\langle v_e \rangle = 1.6 \langle u_1 \rangle$ as the data. The proposed velocity model implies that
the PDF of pore diameters can be characterized by a Weibull
distribution with $\alpha = 2.5$. Figure~\ref{fig:aPDF} shows the
grain size PDF $p_d(d)$ of grain diameters and the PDF of pore
diameters $p_a(a)$ estimated from the above velocity model. The
distribution of pore diameters is much more variable than the grain size
distribution, specifically toward low values. This is intuitive
because the pore size is related to the distance between grains in the
packings, which can be much smaller than the grain size, see also
Figure~\ref{fig:2d3d}. 

From the PDF of pore diameters, we can estimate the PDF of mean
velocities $v_m = v_p/2$, which is
\begin{align}
p_m(v) = \int d a \frac{b a^2 \ell_0}{\langle V_p \rangle}
  \delta[v_0  (a/a_0)^2/2 - v] p_a(a),
\end{align}
By setting again $\langle V_p \rangle \propto a_0^2 \ell_0$
and using the Weibull distribution~\eqref{weibull}, we obtain
\begin{align}
p_m(v) \propto \frac{1}{v_0} 
\int d a \delta(a^2 - 2 v/v_0) \alpha a^{\alpha+1}
  \exp[-a^\alpha]. 
\end{align}
Evaluating the Dirac delta and normalising gives 
\begin{align}
\label{pmv}
p_m(v) = \frac{\alpha}{v_0 \Gamma(1+2/\alpha)} \left(\frac{2 v}{v_0}
  \right)^{\frac{\alpha}{2} }
  \exp\left[-\left(\frac{2 v}{v_0}\right)^{\alpha/2}\right].
\end{align}
We thus obtain for the average of $v_m$
\begin{align}
\langle v_m \rangle = v_0 \frac{\Gamma(1 + 4/\alpha)}{2 \Gamma(1 +
  2/\alpha)} = v_0 \frac{\Gamma(4/\alpha)}{\Gamma(2/\alpha)} \equiv
  \langle v_e \rangle
\end{align}
Expression~\eqref{pmv} implies that $p_m(v) \propto v^{a -1}$
with $a = \alpha/2+1$. The PDF of the mean velocities is
characterised by lower weights than the Eulerian PDF on both low and
high velocities. This is illustrated in Figure~\ref{fig:vpdf}. We
approximate~\eqref{pmv} in the following by the $\Gamma$--distribution
\begin{align}
\label{pmgamma}
p_m(v) = \frac{1}{v_\gamma\Gamma(a)}
  \left(\frac{v}{v_\gamma}\right)^{a-1} \exp\left(- \frac{v}{v_\gamma}
  \right),
\end{align}
where $a = \alpha/2 +1$ and $v_\gamma = \langle v_e \rangle/a$. The
quality of this approximation is shown in Figure~\ref{fig:vpdf}. 
As we will see in the following, the PDF of mean velocities plays an
important role for the interpretation of solute dispersion under
finite P\'eclet numbers.  
\section{Pore-Scale Transport Mechanisms and Upscaling\label{sec:CTRW}}
We study the pore scale transport mechanisms in the
light of heterogeneous advection and diffusion in the pore space. 
As discussed above, pore scale transport mechanisms have been
characterised by advection-dispersion models in terms of
the average pore velocity $\langle u_1 \rangle$ and effective
dispersion coefficients $\mathcal
D$~\cite[e..g,][]{icardi2014pore,scheven2007intrinsic,Bijeljic2004},
and dual domain models~\cite[][]{liu2012applicability,gjetvaj2015} to
capture the impact of diffusion into the immobile porosity and low
velocity regions. The impact of advective heterogeneity 
has been studied in the framework of continuous time and time domain random
walks~\cite[][]{Bijeljic2006,LeBorgneWRR2011, bijeljic2011signature,
  DeAnna2013,Kang2014,gjetvaj2015,Holzner2015}. 
Here we analyse in detail the interplay of heterogeneous advection from the intra to the inter-pore
scale and diffusion in order to identify and quantify the dominant
transport mechanisms. 

\begin{figure}
\begin{center}\includegraphics[width =
  .7\textwidth]{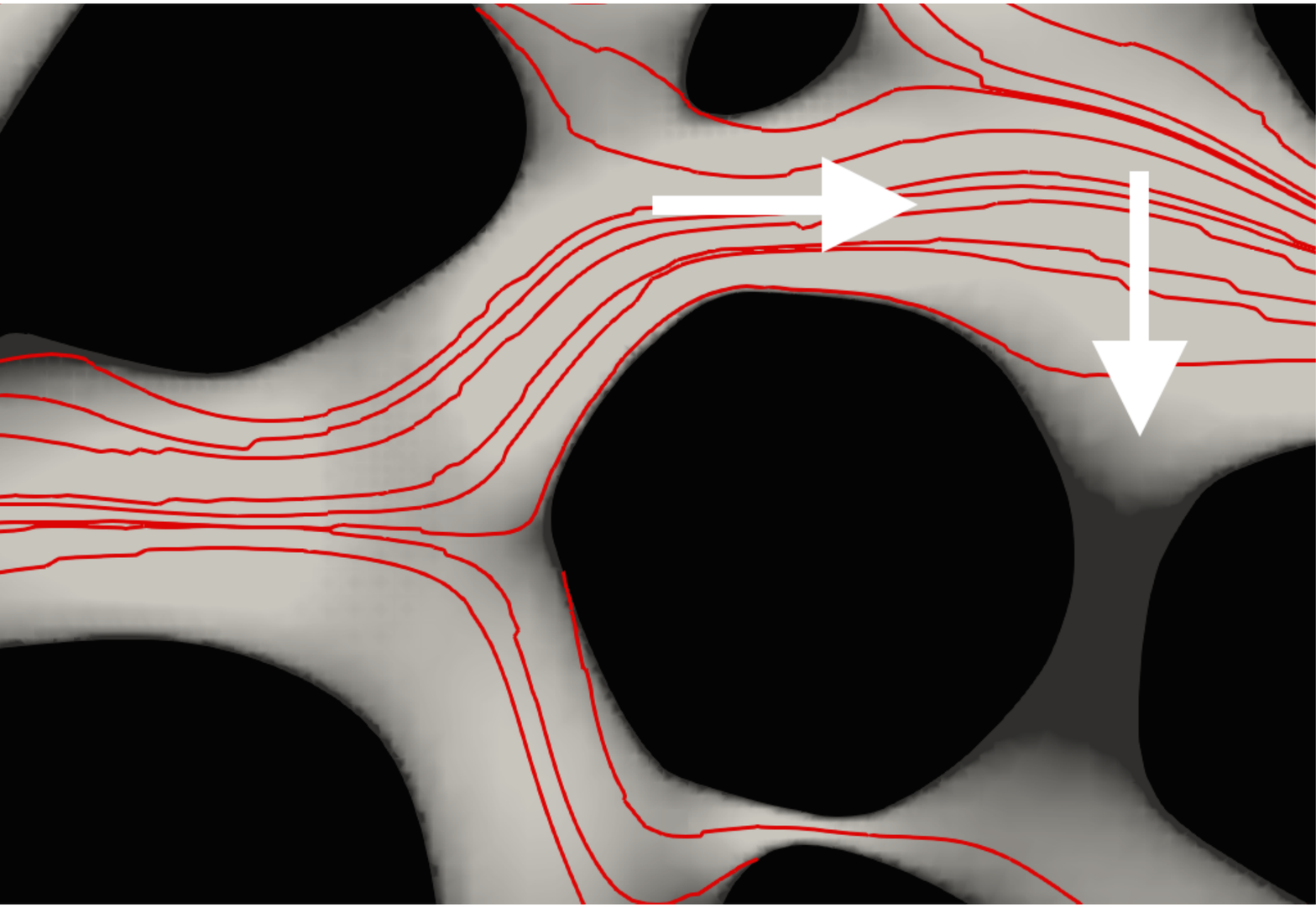}
\end{center}
\caption{Schematic of transitions in the time-domain random walk
  model~\eqref{tdrw}. The thin red lines represent streamlines of the flow
  in a cross-section of the porous medium. The shading of the
  background denotes the absolute value of
  the flow velocity, which decreases from light to dark, black areas
  denote the grains. The horizontal arrow denotes
  particle transitions over the length of a pore
  in streamwise direction, the vertical arrow indicates particle  
  transitions across streamlines into low velocity zones in the wake of
  the solid grains. Transitions along streamlines are due to mean
  advection in the pore and diffusion. Transitions across the
  streamlines are due to diffusion. Particles that enter the immobile
  zones remain there for the residence time $\tau_f$ before they are
  released back into the mobile portion of the medium.    
  \label{fig:schematic}}
\end{figure}

To this end, we model solute transport in a particle-based framework
through spatial transitions over the characteristic pore length $\ell
\sim \ell_0$ with a (variable) duration $\tau$ \citep{saffman1959theory}. 
In this picture, the position $x_n$ and time $t_n$ of a solute particle along the
mean flow direction are given after $n$ steps as
\begin{align}
\label{tdrw}
x_{n+1} = x_n + \xi_n, && t_{n+1} = t_n + \tau_n.
\end{align}
The spatial step $\xi_n$ may be downstream or upstream depending on
the local advection and diffusion conditions. The transition length is given by
$|\xi_n| = \ell/\chi$. Particles make transitions along their
trajectory of length $\ell$. The random walk~\eqref{tdrw}, however, describes
transport along the mean flow direction. Thus, the transition length
is corrected by the advective tortuosity $\chi$, which accounts for the
fact that particle trajectories through the pore space are not
straight. The transition time $\tau_n$ depends on the pore-scale advection and
diffusion mechanisms as discussed in detail below, and illustrated
schematically in Figure~\ref{fig:schematic}. 
 We assume that the advection and diffusion
properties at subsequent positions are statistically independent,
which implies that the $\tau_n$ are identical independent random variables.  
This is a reasonable assumption because flow and particle velocities
are only weakly correlated between pores~\cite[][]{saffman1959theory, LeBorgneWRR2011,
  bijeljic2011signature, DeAnna2013, lester2013chaotic, Kang2014,
  Holzner2015}. 

Equation~\eqref{tdrw} describes a time-domain random walk
(TDRW)~\cite[][]{Cvetkovic1991, Delay:et:al:2005, Painter2005,russian2016,noetinger2016}, or
equivalently a continuous time random walk (CTRW)~\cite[][]{MW1965,
  SL73.1, Berkowitz2006}. It is characterised by the joint
PDF of transition length $\xi$ and time $\tau$ denoted by
$\psi(x,t)$, which encodes the pore-scale transport mechanisms. The
marginal PDF of transition times is defined by 
\begin{align}
\label{psit}
\psi(t) = \int dx \psi(x,t). 
\end{align}
The evolution of the macroscale solute
concentration in this framework is given by the generalised
Master equation~\cite[][]{KMS73, Berkowitz2006, Comolli2016} 
\begin{align}
\label{gme}
\frac{\partial \overline c(x,t)}{\partial t} = \int dx' \int\limits_0^t
  dt' \mathcal K(x-x',t-t') \left[\overline c(x',t') - \overline c(x,t')
  \right]. 
\end{align}
It expresses the fact that the change of solute concentration at a
given position depends on the transport history and in this sense
accounts for incomplete mixing, or non-uniqueness of concentration on
the support scale. The macroscale concentration at a given position is
composed of solute particles with different transport histories, which is
quantified by the distribution of transition times. The memory
kernel $\mathcal K(x,t)$ denotes the probability per time of a
particle transition of length $x - x'$ during time $t - t'$. It
relates solute fluxes at earlier times $t' < t$ to
concentration changes at time $t$. Under Markovian conditions, the
change of the solute concentration is only determined by the local in
time solute fluxes. The memory kernel is given in terms of the
distribution of transition length and times. It is defined through its
Laplace transform as
\begin{align}
\label{k}
\mathcal K^\ast(x,\lambda) = \frac{\lambda \psi^\ast(x,\lambda)}{1-
  \psi^\ast(\lambda)}. 
\end{align}
The Laplace transform is defined in~\cite{AS1972}, Laplace transformed
quantities are marked by an asterisk, the Laplace variable is denoted
by $\lambda$. If the kernel $\mathcal K(\vx,t)$ is short-range in $x$,
this means it decays sufficiently fast for $|x|$ larger than a
characteristic length scale,~\eqref{gme} can be localised in
space~\cite[][]{Berkowitz2002a, DCSB2004},
\begin{align}
\label{nlade}
\frac{\partial \overline c(x,t)}{\partial t} + 
\int\limits_0^t dt' \left[\nu(t - t') \frac{\partial \overline c(x,t')}{\partial x} - \kappa(t - t') 
\frac{\partial^2 \overline c(x,t')}{\partial x^2} \right] = 0. 
\end{align}
The drift and diffusion kernels are defined by 
\begin{subequations}
\label{kdef}
\begin{align}
\label{nu}
\nu(t) &= \int dx x \mathcal K(x,t)
\\
\label{kappa}
\kappa(t) &= \frac{1}{2} \int dx x^2 \mathcal K(x,t). 
\end{align}
\end{subequations}

In the following, we solve for the macroscale transport
behaviour using random walk particle tracking simulations based
on~\eqref{tdrw}. The solute arrival time at a position $x$ is
thus given by~\cite[][]{Comolli2016} 
\begin{align}
\tau_a(x) = t_{n_a(x)}, && n_a(x) = \inf(n|x_n \geq x), 
\end{align}
where $n_a(x)$ is the number of steps a particle requires to arrive at
the outlet. The distribution of arrival times, denoted by $p_a(t,x)$
is defined by 
\begin{align}
p_a(t,x) = \langle \delta[t - \tau_a(x)]\rangle. 
\end{align}
The complementary cumulative breakthrough curve, or in the following
only breakthrough curve, is thus given by 
\begin{align}
\label{ftdrw}
f(t,x) = \int\limits_t^\infty dt' p_a(t',x) = \langle H[\tau_a(x) - t] \rangle, 
\end{align}
where $H(t)$ is the Heaviside step function. Expression~\eqref{ftdrw}
is the probability that the arrival time is larger than $t$. In the following, we denote the breakthrough curve at the
outlet at $x = L$ by $f(t) \equiv f(t,x=L)$. 

Equation~\eqref{tdrw} for the particle motion and its
counterpart~\eqref{gme} for the particle distribution describe the
non-local evolution of macroscale concentration due to local
physical non-equilibrium, but are not
predictive as long as $\psi(x,t)$ is not known. However, the TDRW
approach provides a framework to identify and quantify the dominant pore-scale transport
mechanisms in terms of the distribution of the Eulerian flow velocity
and the diffusion properties in the medium. Thus, in the following we
investigate the structure of $\psi(x,t)$ in the light of the dominant
pore-scale transport mechanisms and their manifestations on large
scale transport. 

Figure~\ref{fig:schematic} show a schematic of
possible particle transitions. Solute particles move along the
pores due to advection-diffusion and may access low velocity regions in
the wake of the grains by diffusion only. 
We first consider the impact of heterogeneous advection and
diffusion along streamlines on the behaviour of the arrival time
distribution. Then we study the influence of diffusion across
streamlines into low velocity zones and its quantification. Finally, we
consider the combined impact of advective heterogeneity and
diffusion along and across pores on large scale transport. 
\subsection{Heterogeneous advection and diffusion}
We analyse here the role of heterogeneous advection and diffusion
along the streamlines within and between pores. To this end, we first
consider purely advective particle motion and the resulting large
scale transport behaviour in terms of the particle arrival time
distributions. Then, we analyse the effect of diffusion first on the
intra-pore and then on the inter-pore particle motion. 
\subsubsection{Pure advection}
For infinite $Pe$, this means, purely advective
transport,~\eqref{tdrw} becomes 
\begin{align}
\label{tdrwa}
x_{n+1} = x_n + \frac{\ell}{\chi}, && t_{n+1} = t_n + \frac{\ell}{v_n}
\end{align}
Particles move only forward along streamlines and the transition times
$\tau_n = \ell/v_n$ denote the advection time over the distance $\ell$
by the velocity magnitude $v_n$. The magnitudes $v_n$ of particle
velocities are distributed according to the
stream-wise velocity PDF $p_s(v)$ defined as follows. First we note
that the TDRW~\eqref{tdrw} samples particle velocities equidistantly along trajectories, while
the Eulerian velocity PDF is sampled volumetrically in space. For divergence-free flow, $p_s(v)$ is 
 related to the Eulerian velocity PDF through flux
 weighting~\cite[][]{saffman1959theory, Dentzctrwprf2016, comolli2017}
\begin{align}
\label{pflux}
p_s(v) = \frac{v p_e(v)}{\langle v_e \rangle}.
\end{align}
This relation can be understood qualitatively by noting that 
low flow velocities occupy wider streamtubes than high velocities
because of fluid volume conservation. Thus, volumetric sampling
emphasises low flow velocities. Equidistant velocity sampling
along streamlines on the other hand, weighs high and low velocities
equally. This difference in sampling between the Eulerian and
streamwise velocity distributions is compensated in~\eqref{pflux}
through flux weighting. 

With this relation, the PDF of purely advective transition times
denoted here by $\psi_a(t)$ can be directly
related to the distribution $\psi_a(t)$ of advective transition times as
\begin{align}
\label{psia}
\psi_a(t) = \frac{\ell}{t^3 \langle v_e \rangle} p_e(\ell/t). 
\end{align}
The Eulerian PDF $p_e(v)$ goes
towards a constant at small velocities and is cut-off as a stretched
exponential at large velocities, see Figure~\ref{fig:vpdf}
and~\eqref{pe}. For the numerical implementation of the TDRW, we
approximate the Eulerian velocity PDF~\eqref{pe} by an exponential distribution
characterised by the same mean velocity. Specifically, the characteristic that
determines the late time behaviour of the breakthrough curve is the
behaviour of $p_e(v)$ at small velocities $v \ll \langle v_e \rangle$,
for which it is constant. Thus, from~\eqref{psia}, we obtain that  the PDF of transition
times behaves as $\psi_a(t) \propto t^{-1-b}$ with $b = 2$ for
times $t \gg \tau_u$. Thus, $\psi_a(t)$ is an asymptotically stable
distribution. 
\begin{figure}
a.\includegraphics[width = .45\textwidth]{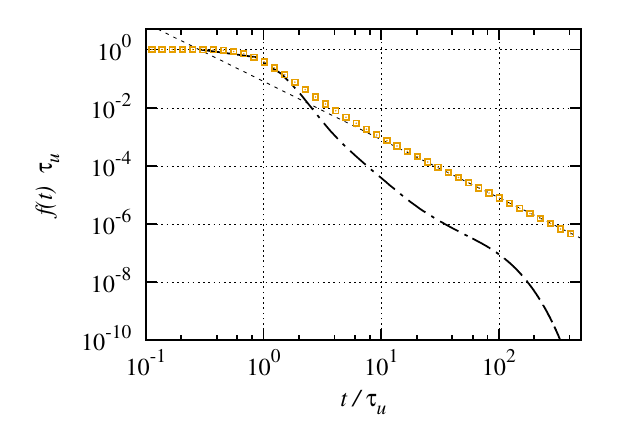}
b.\includegraphics[width = .45\textwidth]{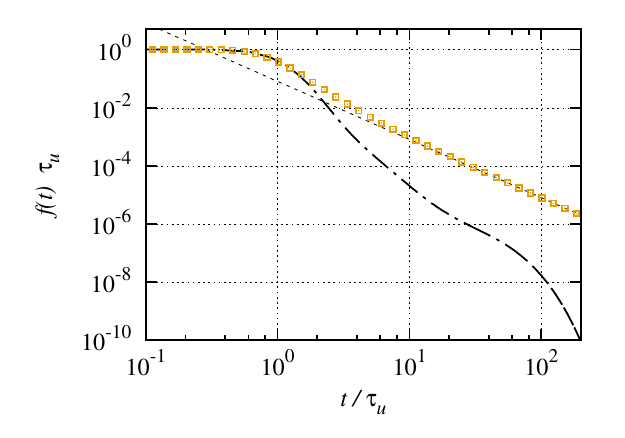}

c.\includegraphics[width = .45\textwidth]{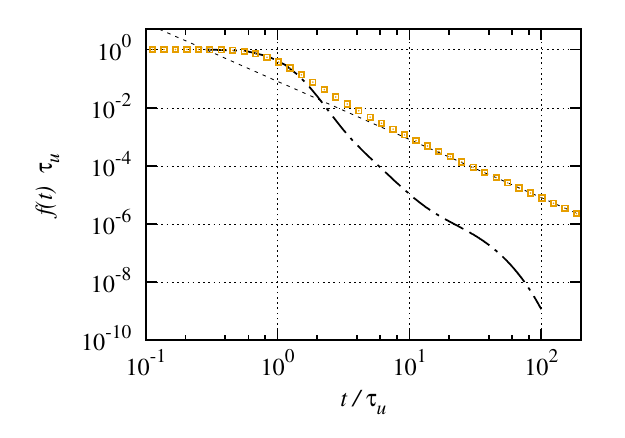}
d.\includegraphics[width = .45\textwidth]{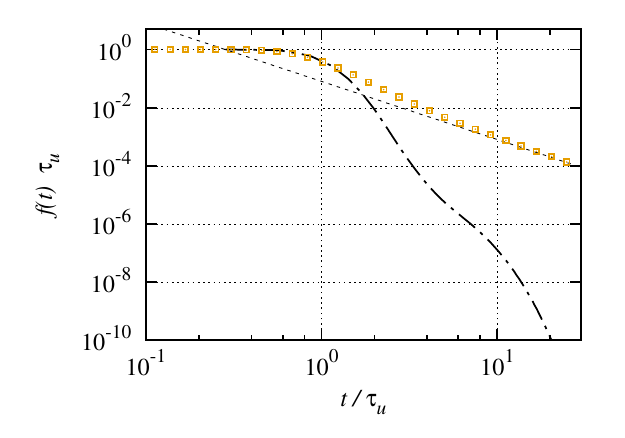}
\caption{Breakthrough curves for $Pe = 10^3$, $5 \cdot 10^2$, $3 \cdot
  10^2$, and $30$. The symbols denote the breakthrough curves obtained
  from the TDRW~\eqref{tdrwa} based on purely
  advective transport, $Pe = \infty$, for the transition length
  $\ell \approx  0.84 \ell_p$. The dashed lines show the predicted
  $t^{-2}$--scaling. The TDRW simulations use $10^9$ particles.    
  \label{fig:btc2}}
\end{figure}

The breakthrough, or arrival times at
the outlet are given by 
\begin{align}
\tau_a = \sum\limits_{i = 1}^{n_a} \tau_i, 
\end{align}
where the number of steps $n_a = \lceil L \chi /\ell \rceil$,
where $\lceil \cdot \rceil$ is the ceiling function. As
$\psi_a(t)$ is stable, the PDF of arrival times, as a sum of stable
random variables, has the same long time behaviour as $p_f(t,x) \propto
t^{-1-b}$. Thus, the breakthrough curve is $f(t,x)
\propto t^{-b}$, see~\eqref{ftdrw}. This means here that the breakthrough curve is
predicted to behave as $f(t) \propto
t^{-2}$. Figure~\ref{fig:btc2} compares the numerically obtained
breakthrough curves for different P\'eclet numbers to the predictions
of~\eqref{tdrwa} parameterised by the advective transition time
PDF~\eqref{psia}. The purely advective breakthrough curve is of course
independent of $Pe$ and shows the $t^{-2}$-decay as predicted. Also
here, the onset and initial decay of the breakthrough curves are well
captured, but the purely advective model fails to represent the tails
of the pore-scale data. It overestimates the tailing, while the ADE
model underestimates it. 

For all $Pe$, we observe two distinct tailing
behaviours. An intermediate tail whose decay is slower than the
exponential tail predicted by the ADE, and a flatter long-time tail
that is cut-off exponentially. The slope of the intermediate tail and
the onset and final decay of the late-time tail clearly depend on the
P\'eclet number and thus on diffusion. In the following, we study
these mechanisms in more detail. 
\subsubsection{Advection under diffusion\label{sec:hetadv}}
Diffusion acts on how particles experience the flow velocities in two ways. 
Firstly, within a pore diffusion enables particles to sample the flow profile across
the pore~\cite[][]{saffman1959theory, haber1988lagrangian, Dentz:Carrera:2007}. This
implies that the effective velocity experienced by a particle along a
pore tends toward the mean velocity $v_m$. The velocity extremes are
attenuated by diffusion. It is these extremes, particularly the low
velocities close to the pore walls, which cause the $t^{-2}$-tails
predicted by the purely advective model. If the diffusion time
across the pore is smaller than the characteristic advection time
along the pore, particles move effectively with the mean velocity
$v_m$~\cite[][]{saffman1959theory}. This is a good approximation here
because the estimated distribution of pore diameters is much broader
than the distribution of pore lengths, and tailed towards small values
as shown in Figure~\ref{fig:aPDF}. This means that the transport
relevant velocity distribution is the PDF of mean pore
velocities~\eqref{pmv} rather than the PDF~\eqref{pe} of
point-wise velocities. Note that the mean pore velocities are still sampled volumetrically,
while the particle velocities are sampled spatially along
streamlines. Thus, as outlined above, the distribution of
particle velocities is given by flux-weighting of $p_m(v)$ as
\begin{align}
p_s(v) = \frac{v p_m(v)}{\langle v_e \rangle}. 
\end{align}
Recall that $\langle v_m \rangle = \langle v_e \rangle$. 

Secondly, the transition time over the pore length is the
result of  advection by the mean pore velocity and diffusion. For
example, for pores whose advection time $\ell/v_m$ is larger than the diffusion
time $\ell^2/2D$ along the pore, the particle transition time is dominated by
diffusion. The transition time is limited by the characteristic
diffusion time~\cite[][]{saffman1959theory}. The distribution of
advective-diffusive transit times over a given distance can be seen as
a first passage problem~\cite[][]{Redner} characterised by $v_m$ and
the diffusion coefficient $D$. Here, these mechanisms are captured by
quantifying the transition time as an exponentially distributed random variable whose mean is
given in terms of the harmonic mean between the advection and the
diffusion times over the length $\ell$~\cite[][]{Delay:et:al:2005,noetinger2016,russian2016}
\begin{align}
\label{psi0}
\psi_0(t|v) = \frac{\exp(-t/\tau_v)}{\tau_v}, && \tau_v = \frac{\ell/v}{1 +  \frac{2 D}{v \ell} }
\end{align}
The latter is an approximation for the true first-passage time
PDF and can be derived from a finite volume discretisation of the advection-diffusion
equation~\cite[][]{russian2016}. Furthermore, depending on the
relative strength of diffusion and advection within the pore,
particles may move up or downstream. The probability $w_u(v)$
for an upstream transition is given by~\cite[][]{russian2016} 
\begin{align}
\label{wu}
w_u(v) = \frac{D \tau_v }{\ell^2} 
\end{align}
and for downstream motion correspondingly $w_d(v) = 1 - w_u(v)$. Thus,
the PDF $\Lambda(x|v)$ of transition length for a given
velocity $v$ reads as
\begin{align}
\label{Lambda}
\Lambda(x|v) = \left[w_d(v) \delta(x - \ell/\chi) + w_u(v) \delta(x +
  \ell/\chi)\right]. 
\end{align}
The global joint PDF of transition lengths and times then is given by 
\begin{align}
\label{psim}
\psi_m(x,t) = \int\limits_0^\infty dv \frac{v p_m(v)}{\langle v_e
  \rangle} \Lambda(x|v) \psi_0(t|v).   
\end{align}
The transition time PDF is obtained by marginalisation according
to~\eqref{psit},
\begin{align}
\label{psimt}
\psi_m(t) = \int\limits_0^\infty dv \frac{v p_m(v)}{\langle v_e
  \rangle}  \psi_0(t|v).   
\end{align}

Using these expressions, we obtain for the drift and diffusion
kernels~\eqref{nu} and~\eqref{kappa} the following Laplace space expressions,
\begin{subequations}
\label{kernels}
\begin{align}
\nu_m^\ast(\lambda) &= \int\limits_0^\infty dv \frac{v^2 p_m(v)}{\chi \langle v_e \rangle} \frac{1 - \psi_0^\ast(\lambda|v)}{1 - \psi_m^\ast(\lambda)}
\\
\kappa_m^\ast(\lambda) &= \int\limits_0^\infty dv \frac{ (D + v \ell/2) v p_m(v)}{\chi^2 \langle v_e \rangle} \frac{1 - \psi_0^\ast(\lambda|v)}{1 - \psi_m^\ast(\lambda)},
\end{align}
\end{subequations}
see Appendix~\ref{app:kernels}. 

\begin{figure}
a.\includegraphics[width = .45\textwidth]{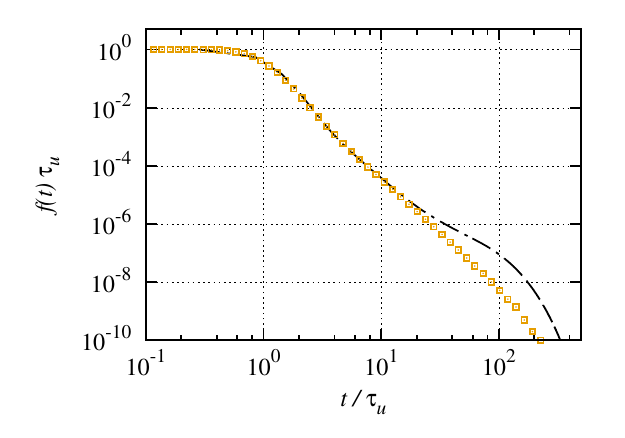}
b.\includegraphics[width = .45\textwidth]{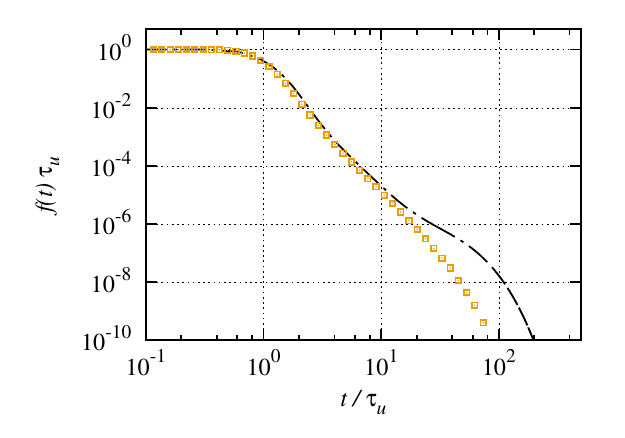}

c.\includegraphics[width = .45\textwidth]{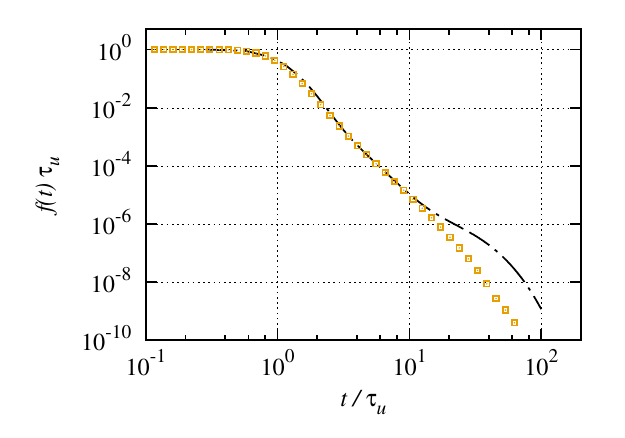}
d.\includegraphics[width = .45\textwidth]{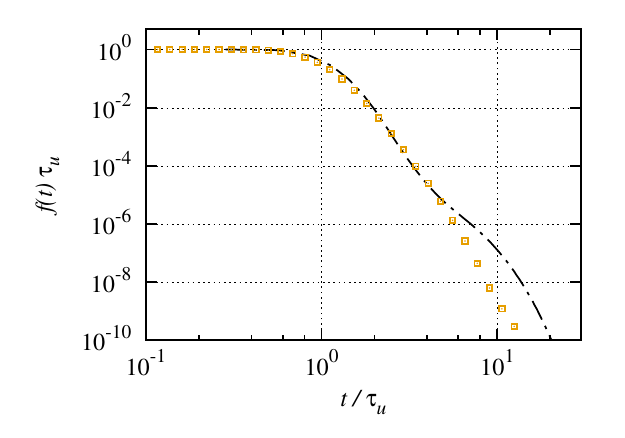}
\caption{Breakthrough curves for $Pe = 10^3$, $3 \cdot 10^2$, $30$ and
  $3$. The dash-dotted lines indicate the numerical data, the symbols
  the TDRW model for $\ell \approx \ell_p$ for $Pe = 10^3$, $\ell = 0.77
\ell_p$ for $Pe = 5 \cdot 10^2$, and $Pe = 3 \cdot 10^2$ and $\ell =
0.7 \ell_p$ for $Pe = 30$. 
  \label{fig:btc3}}
\end{figure}

We solve this TDRW model for the breakthrough curves using random walk
particle tracking simulations. 
Figure~\ref{fig:btc3} shows the results of this TDRW characterized
by~\eqref{psim} compared to the direct numerical simulations of the pore scale flow and transport
problem. Note that the TDRW is parameterized by the PDF of mean pore
velocities and the P\'eclet number. For computational convenience, here and in the following, 
we approximate the PDF~\eqref{pmv} of mean pore velocities
by~\eqref{pmgamma}. The TDRW model captures correctly the
intermediate tailing of the breakthrough curves at all P\'eclet
numbers, which thus can be attributed to both heterogeneity in the
mean pore velocity and diffusion. 

Note that the PDF of transition
times~\eqref{psimt} is not heavy tailed. The maximum transition time
is given by $\ell^2/2D$, which determines the cut-off
of~\eqref{psimt}. For smaller times, it behaves as $\psi_m(t)
\propto t^{-3-\alpha/2}$, which can be seen by inserting~\eqref{pmv} into~\eqref{psimt}
and rescaling of the integration variable. Thus, according to the
central limit theorem, breakthrough curves at large distances from the
inlet are expected to converge to a Fickian limit. In this sense, the
tailing behavior in the intermediate regime is pre-asymptotic and can be
attributed to the distance from the inlet or sample size. At times longer than the
characteristic diffusion time over the pore, the TDRW breakthrough
curves are cut-off and do not capture the second
tailing regime characterized by first a flattening and then
exponential cut-off. In the following, we discuss the origin of the
second tailing regime in the light of trapping due to particle
transitions across streamlines into low-velocity pores.  
\subsection{Solute trapping and release\label{sec:mrmt}}
In the previous section, we have focused on particle motion in
direction of the streamlines by the mean pore velocity as illustrated
schematically by the horizontal arrow in
Figure~\ref{fig:schematic}. Alternatively, particles
may make diffusive transitions across streamlines 
into the wake of a solid grain, indicated by the vertical arrow in
Figure~\ref{fig:schematic}. In these regions, the flow velocity is
close to zero and diffusion is the dominant transport
mechanisms. Thus, we denote these regions in the following as immobile
zones. 
\begin{figure}
a.\includegraphics[width = .45\textwidth]{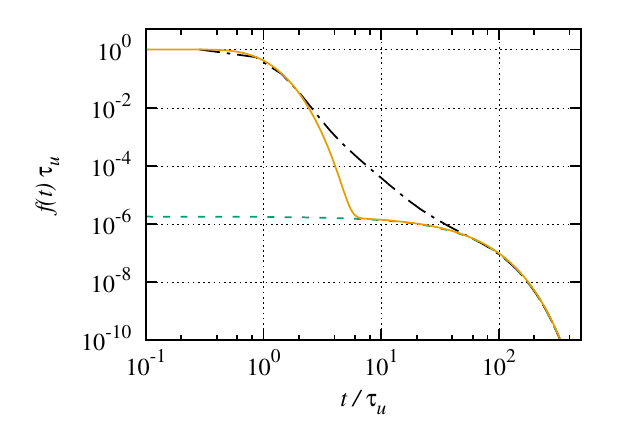}
b.\includegraphics[width = .45\textwidth]{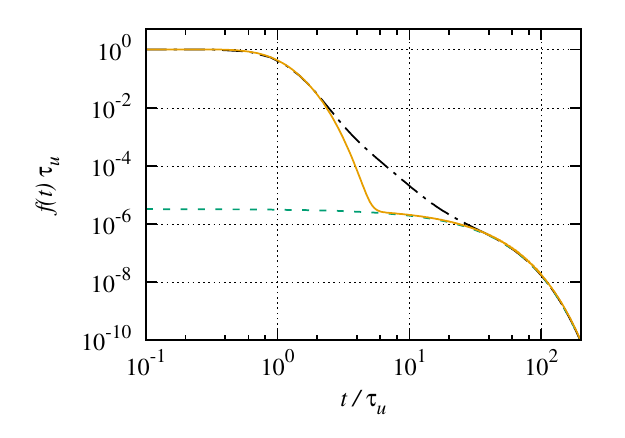}

c.\includegraphics[width = .45\textwidth]{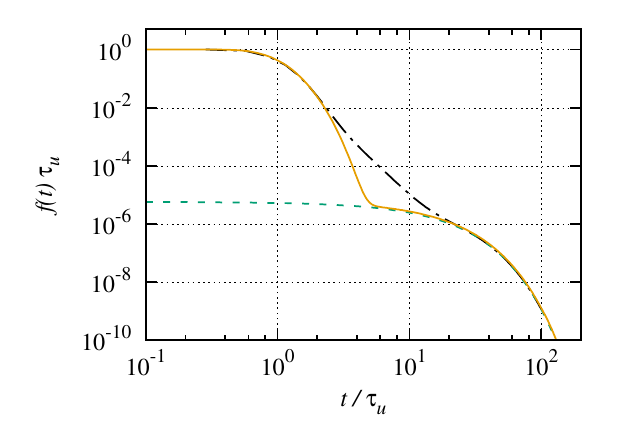}
d.\includegraphics[width = .45\textwidth]{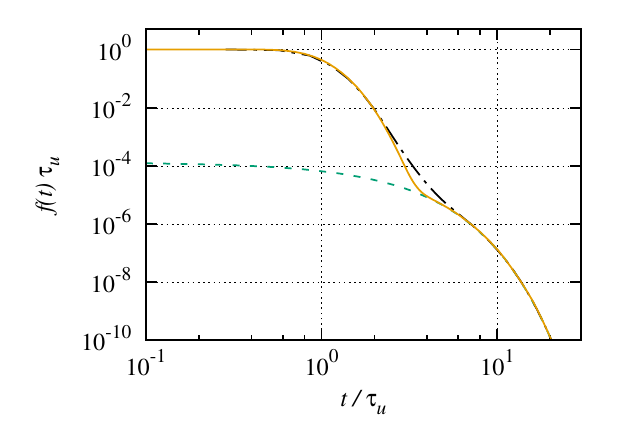}
\caption{Breakthrough curves for $Pe = 10^3$, $3 \cdot 10^2$, $30$ and
  $3$. The dash-dotted lines indicate the numerical data, the circles indicate the exponential estimation of the tailing
behaviour, from which we obtain the values for $\langle \tau_f
\rangle$ and $\gamma$ displayed in Figure~\ref{fig:PeTau} below.\label{fig:btc4}}
\end{figure}

We first focus solely on the impact of cross-diffusion into low
velocity zones and represent heterogeneity of the pore velocities in
terms of the average velocity $\langle v_e \rangle$ and the hydrodynamic dispersion
coefficients $\mathcal D$ determined above. This means that we assume
that advective heterogeneity can be homogenised by
advection-dispersion. Below in Section~\ref{sec:aht}, we combine the
effects of heterogeneous advection and diffusion into immobile zones. As we assume
that mobile transport is homogenised, there is no characteristic
transition length scale in contrast to the previous sections, where the
pore-length $\ell$ marks the correlation scale of the stream-wise
velocities. Velocity fluctuations are quantified by hydrodynamic dispersion. Thus, in terms
of the advective-diffusive TDRW introduced in the previous section,
this means that particles perform spatial transitions over a distance
$\Delta s \ll L$, which now merely represents a discretisation of
space~\cite[][]{russian2016}. Particle transitions in space then are
characterized by~\eqref{wu} and~\eqref{Lambda}
with the substitutions 
\begin{align}
\label{subs}
\ell \to \Delta s, && v \to \langle v_e \rangle, && D \to \mathcal D. 
\end{align}
%
%
Particle transitions in time now are composed
of a mobile time $\tau_0$, which represents advective-dispersive
motion in mean flow direction and the time of trapping in immobile
zones as $\tau = \tau_0 + \tau_{im}(\tau_0)$. The mobile time is
distributed according to~\eqref{psi0} with the
substitutions~\eqref{subs}. The immobile time $\tau_{im}$ is
determined based on the assumption that immobile zones are uniformly
distributed in the medium. As the characteristic time scale
between trapping events is given by the characteristic cross-diffusion
time, trapping events can be assumed to occur at a constant rate
$\gamma$. This means that the number of trapping events $n_{\tau_0}$ 
that occur during a time $\tau_0$ are Poisson-distributed,
\begin{align}
\label{poisson}
p_n(n|\tau_0) = \frac{(\gamma \tau_0)^n \exp(-\gamma \tau_0)}{n!},  
\end{align}
Thus, the immobile time $\tau_{im}$ during a mobile transition of
duration $\tau_0$ is given by the sum over the $n_{\tau_0}$
individual trapping times times $\tau_f$,
\begin{align}
\label{tauim}
\tau_{im}(\tau_0) = \sum\limits_{k = 1}^{n_{\tau_0}} \tau_{f,k}. 
\end{align}
It is a compound Poisson process~\cite[][]{Feller1}. 
The trapping times $\tau_f$ are independent identically distributed
according to $p_f(t)$, which is determined below. The compound Poisson process~\eqref{tauim}
is characterized by the  PDF $\psi_f(t|\tau_0)$, whose Laplace
transform is given by~\cite[][]{Margolin:et:al:2003}
\begin{align}
\label{psicp}
\psi_f^\ast(\lambda|\tau_0) = \exp\left(- \left\{\lambda + \gamma\left[1 - p_f^\ast(\lambda) \right] \right\} \tau_0 \right). 
\end{align}

In order to quantify the impact of cross-diffusion on macroscale
transport, we need to determine the distribution of trapping times
$\tau_f$ and the trapping rate $\gamma$. To
this end, we first note that the total concentration can be decomposed
into the concentration $c_m(x,t)$ of the solute in the mobile and
$c_{im}(x,t)$ in the immobile regions, $\overline c(x,t) = c_m(x,t) +
c_{im}(x,t)$. Note that here both mobile and immobile concentrations
refer to the same support volume.  
The mobile concentration in this TDRW framework is described by the non-local
advection-diffusion equation~\cite[][]{Margolin:et:al:2003, BensonMeer2009, russian2016,
  Comolli2016}
\begin{align}\label{mrmt}
\frac{\partial c_m(x,t)}{\partial t}  + \langle u_1 \rangle \frac{\partial
  c_m(x,t)}{\partial x} - \mathcal D \frac{\partial^2 c_m(x,t)}{\partial
  x^2} = - \frac{\partial  c_{im}(x,t)}{\partial t}. 
\end{align}
The immobile concentration $c_{im}(x,t)$ is given by 
\begin{align}
\label{cim}
c_{im}(x,t) = \int\limits_0^t dt' \left[\int\limits_{t -t'}^\infty dt''
  p_f(t'') \right] \gamma c_m(x,t').
\end{align}
The expression in the square brackets denotes the probability that the
trapping time is larger than $t - t'$. This means, the immobile
concentration is given by the probability per time that a particle enters in
the immobile zone expressed by $\gamma c_m(x,t)$ times the probability
that it stays there for a time larger than the observation
time. The ratio $\beta$ of total mass in the immobile and mobile zones at equilibrium is obtained 
from~\eqref{cim} as 
\begin{align}
\beta = \gamma \langle \tau_f \rangle. 
\end{align}
Note that~\eqref{mrmt} describes solute transport in the multirate 
mass transfer approach~\cite[][]{Haggerty1995, Carrera1998, Comolli2016}.  
In this framework, $\beta$ is equal to the ratio between the immobile $\phi_{im}$ and 
mobile volume fraction  $\phi_{m}$. Thus, the trapping rate is given by
\begin{align}
\label{gamma}
\gamma = \frac{\phi_{im}}{\phi_m\langle \tau_f \rangle},
\end{align}
Using the mobile and immobile volume fractions, the total
concentration $\overline c(x,t)$ can be written in terms of $g_m(x,t)
= c_m(x,t)/\phi_m$ and $g_{im}(x,t) = c_{im}(x,t)$, which refer to the
mobile and immobile subvolumes, as $\overline c(x,t) = \phi_m g_m(x,t)
+ \phi_{im} g_{im}(x,t)$.  

\begin{figure}
\includegraphics[width = .45\textwidth]{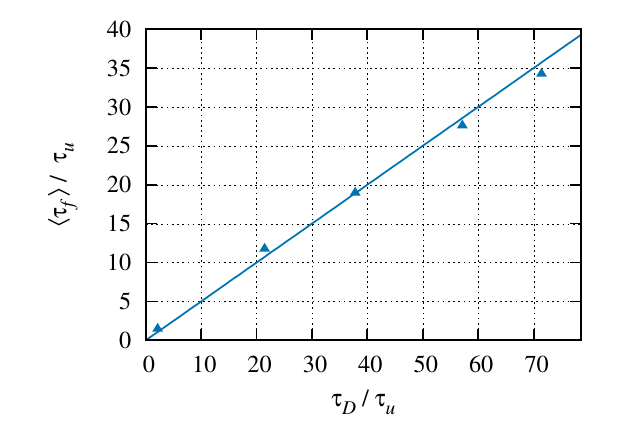}
\includegraphics[width = .45\textwidth]{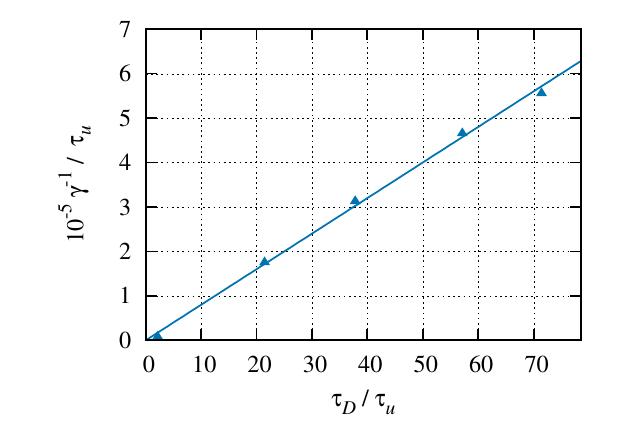}
\caption{(Left panel)  Mean trapping time $\langle \tau_f \rangle$ versus $\tau_D$ estimated from the
data and (solid line) linear estimation with $r = 1/2$. (Right
panel) Inverse trapping rate versus $\tau_D$ and linear estimate with
$\beta = 2.5 \cdot 10^{-4}$.\label{fig:PeTau}}
\end{figure}
From~\eqref{mrmt}, we obtain for the long-time behaviour of the 
breakthrough curves the following behaviour, 
\begin{align}
\label{f:approx}
f(t) = \tau_u \gamma \int\limits_{t}^\infty dt' p_f(t).
\end{align}
for $t \gg \tau_u$. The full solution for $f(t)$ and the derivation of this approximation are detailed in Appendix~\ref{app:fptd}. 
This means, the tailing behaviour is fully determined in terms of the PDF of trapping times. Figure~\ref{fig:btc4} shows that the 
tail of the breakthrough curve can well approximated by an exponential function, which implies that distribution of trapping times can be represented by an exponential PDF. In fact, the distribution of diffusion times in an immobile zone of size $\ell$ can be
obtained through the solution of a diffusion problem as outlined in Appendix~\ref{app:fptd}. The average trapping time is related to the 
characteristic diffusion time $\tau_D = \ell^2/(2D)$ in a domain of length $\ell$. 
The PDF of trapping times for such a problem is well approximated again by the exponential distribution
 as~\cite[][]{Delay2002,DentzGouzeAWR2012}.  
\begin{align}
\label{psif}
p_f(t) = \frac{\exp(-t/\langle \tau_f \rangle)}{\langle \tau_f \rangle}, 
\end{align}
see also Appendix~\ref{app:trap}. The mean trapping time is $\langle \tau_f \rangle = r \tau_D$ with $r \sim 1$. 
The tail of the breakthrough curve is thus given by 
\begin{align}
f(t) = \tau_u \gamma \exp(-t/r\tau_D). 
\end{align}
This means, we can obtain both the trapping rate and the mean trapping time from the tails of the breakthrough 
curves at different P\'eclet numbers. According to our reasoning, both depend on $\tau_D$ as
\begin{align}
\label{taugamma}
\langle \tau_f \rangle = r \tau_D, && \gamma = \frac{\beta}{r \tau_D},
\end{align}
The dependences of $\langle \tau_f \rangle$ and $\gamma^{-1}$ on $\tau_D$ estimated from the numerical 
data are shown in Figure~\ref{fig:PeTau}.  From the data, we obtain $r \approx 1/2$ and $\beta \approx 2.5 \cdot 10^{-4}$, 
which is consistent with the probability of obtaining velocities smaller than $10^{-3} \langle v \rangle$, which we deem immobile.
 Note from Figure~\ref{fig:vpdf} that the PDF of the velocity magnitude deviates from the flat behaviour characteristic for in-pore velocities at $10^{-3}$ and increases towards lower velocities. This increase may be assigned to low velocities in immobile zones. 

Figure~\ref{fig:btc4} compares the breakthrough curves obtained from the solution of~\eqref{mrmt} (see Appendix~\ref{app:fptd}) parameterized with an exponential PDF of trapping times and the average trapping times and trapping rates given by~\eqref{taugamma} as well as the mean velocity $\langle u_1 \rangle$ and hydrodynamic dispersion coefficient $\mathcal D$. It captures well the onset of decay of the breakthrough curves just like the ADE solution discussed above, and the exponential long-time tail. The onset of the exponential tail depends on the trapping rate $\gamma$, 
the exponential cut-off time is given by the diffusion time scale $\tau_D$. 
As expected, this TDRW model does not represent the intermediate tailing due to heterogeneous pore velocities. The quantification of the combined effect of trapping and heterogeneous advection is discussed in the next section. 
\subsection{Advective Heterogeneity and Trapping\label{sec:aht}}
In this section, we integrate heterogeneous pore-scale advection, and diffusion into 
immobile zones into a TDRW approach that quantifies the dominant pore-scale transport 
mechanisms. This approach models mobile particle transition over the length of a pore as 
detailed in Section~\ref{sec:hetadv}. Solute trapping is modelled as discussed in the previous section. 
Specifically, the number of trapping event is given by the Poisson distribution~\eqref{poisson} conditioned 
on the time $\tau_m = \ell/v$ for a mobile transition. Thus, the transition time 
in this integrated TDRW is~\cite[][]{russian2016,Comolli2016} 
\begin{align}
\label{taumrmt}
\tau = \tau_m + \sum\limits_{k = 1}^{n_{\tau_m}} \tau_{f,k},
\end{align}
where $n_{\tau_m}$ is distributed according to the Poisson distribution $p_n(n|\tau_m)$ and $\tau_f$ according to~\eqref{psif}. 
Note that the total trapping time during a mobile transition given by the second term on the right side of~\eqref{taumrmt} is 
a compound Poisson process. The joint PDF $\psi(x,t)$ of transition lengths $\xi$ and time $\tau$ can then be expressed in Laplace space in terms of the Laplace 
transforms of $\psi_m(t)$, the PDF of mobile times, and the trapping time PDF $p_f(t)$ as
\begin{align}
\label{psimrmt}
\psi^\ast(x,\lambda) = \psi^\ast_m(x,\lambda + \gamma [1 - p_f^\ast(\lambda)]),
\end{align}
see Appendix~\ref{app:psi}. The distribution of transition times is
given accordingly by 
\begin{align}
\psi^\ast(\lambda) = \psi^\ast_m(\lambda + \gamma [1 - p_f^\ast(\lambda)]).
\end{align}
With these results, we obtain for the velocity and diffusion
kernels~\eqref{nu} and~\eqref{kappa} the Laplace space expressions
\begin{subequations}
\label{kernelsmrmt}
%
\begin{align}
\nu^\ast(\lambda) &= \frac{\nu_m[\lambda \Phi^\ast(\lambda)]}{\Phi^\ast(\lambda)}
\\
\kappa^\ast(\lambda) &= \frac{\kappa_m[\lambda \Phi^\ast(\lambda)]}{\Phi^\ast(\lambda)}.    
\end{align}
\end{subequations}
where $\Phi^\ast(\lambda) = 1 + \lambda^{-1} \gamma[1 - p_f^\ast(\lambda)]$, 
see Appendix~\ref{app:kernels}. Following the developments in the previous section, 
we define the immobile concentration $c_{im}(x,t)$ in terms of the mobile concentration as in~\eqref{cim}, 
Thus, based on this relation and expressions~\eqref{kernelsmrmt} for the kernels, we derive from~\eqref{nlade}
the following governing equation for the solute concentration in the mobile regions,  
\begin{align}
&\frac{\partial c_m(x,t)}{\partial t} + \frac{\partial}{\partial t} \int\limits_0^t dt' \left[\int\limits_{t -t'}^\infty dt''
  p_f(t'') \right] \gamma c_m(x,t') =
\nonumber\\
&-\int\limits_0^t dt' \int\limits_0^{t-t'} dt''  
\left[\nu_m(t'') \frac{\partial}{\partial x} - 
\kappa_m(t'') \frac{\partial^2 }{\partial x^2} \right]
 \psi_{f}(t-t'-t''|t'') c_m(x,t'),
\label{cadvmrmt3}
\end{align}
see Appendix~\ref{app:advmrmt}. The second term on the left side denotes the change of the mobile solute concentration due to 
mass transfer with the immobile regions. The term on the right side denotes the solute flux in the mobile region during the period of 
time $t''$ for which particles are making mobile transitions. Note that $\psi_{f}(t-t'-t''|t'')$ denotes the probability that the particle is trapped 
during the time $t - t' - t''$ for a given mobile duration $t''$. 
\begin{figure}
\label{fig:btc-fit}
\includegraphics[width = .45\textwidth]{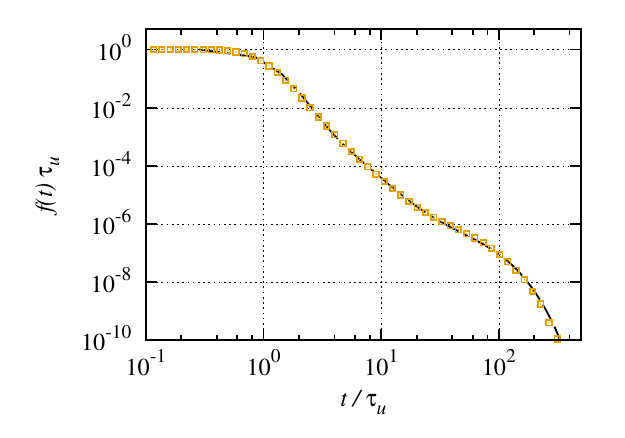}
\includegraphics[width = .45\textwidth]{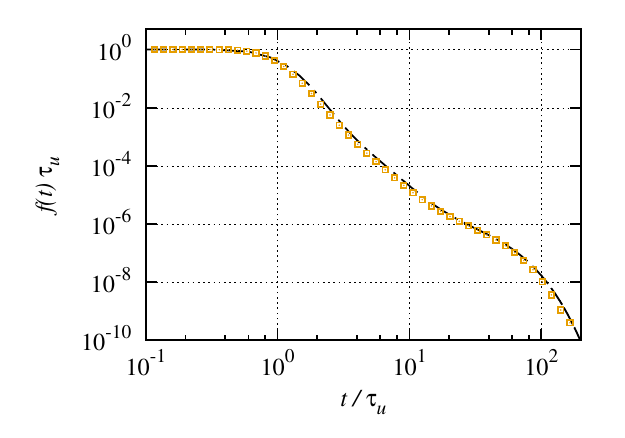}

\includegraphics[width = .45\textwidth]{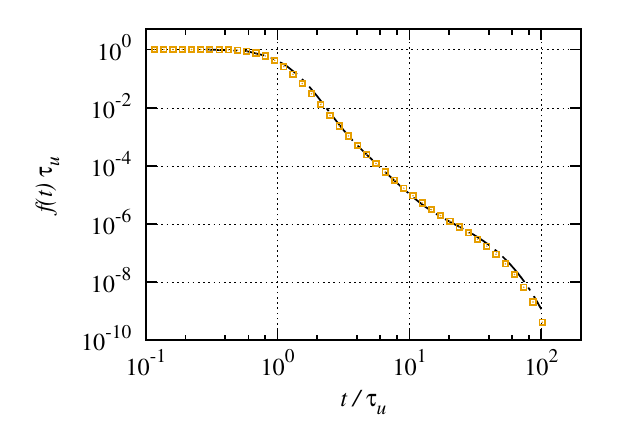}
\includegraphics[width = .45\textwidth]{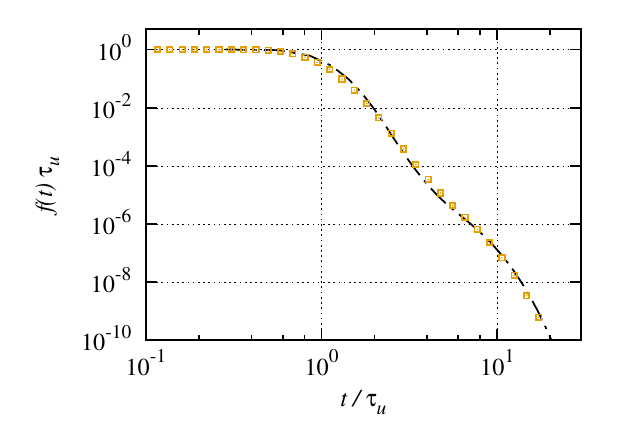}
\caption{Breakthrough curves for $Pe = 10^3$, $5 \cdot 10^2$, $3 \cdot 10^2$ and
  $30$. The dash-dotted line indicate the numerical results.  
The symbols denote the result of the TDRW model including heterogeneous advection and trapping in immobile
zones. The TDRW model uses the parameter values given in the captions
of Figures~\ref{fig:btc3} and~\ref{fig:btc4}. 
}
\end{figure}

We solve this TDRW model for macroscale transport using random walk particle tracking simulations. The transport behaviour is parameterized 
in terms of the PDF of mean pore-velocities, and the trapping rate $\gamma$,~\eqref{gamma}, and the distribution~\eqref{psif} of trapping times discussed in the previous section. Thus, in principle, macroscale transport can be understood in terms of the medium geometry (distribution of pore diameters), hydrodynamics (Poiseuilles's law), and diffusion along and across streamlines into immobile regions.    

Figure~\ref{fig:btc-fit} compares the breakthrough curves from the direct numerical simulations to the behaviour of the TDRW approach. The TDRW 
model and simulation data compare well for all P\'eclet numbers. The intermediate tailing is well captured by the distribution of mean pore-velocities and diffusion in mean flow direction as discussed in Section~\ref{sec:hetadv}. The exponential tail is picked up by the mass transfer model discussed in the previous section. Also, the transitions from the intermediate advective tailing behaviour to the exponential are well represented for all P\'eclet number. The proposed model captures all aspects of the observed solute breakthrough curves based on the identification and quantification of the dominant pore-scale transport mechanisms.







\section{Summary and Conclusions\label{sec:conclusions}}
We have studied and quantified the advection and diffusion mechanisms
of preasymptotic pore-scale transport and their relations to the
statistical medium and flow properties in terms of the pore size
distribution and the distribution of the magnitude of the Eulerian
velocity. Our study uses data from high performance computational
fluid dynamics simulations of flow and solute transport in a synthetic
$3$--dimensional porous medium. The medium properties resemble the characteristics of
sand or bead packs in grain size distribution and porosity. 

We focus on the complementary solute breakthrough curves at the outlet
of the porous medium, to which we refer as breakthrough curve for
simplicity. The advection-dispersion paradigm predicts an exponential decrease of
the breakthrough curve form $1$ towards $0$ at the mean advection time
over the domain length with a width given in terms of the longitudinal
hydrodynamic dispersion coefficient. For all $Pe \gg 1$ we
observe pronounced deviations from this behaviour, which increase with
increasing $Pe$. The breakthrough curves display two non-Fickian
regimes. The intermediate regime is given by a sub-exponential decay,
which can be attributed to advective heterogeneity. The second,
regime is characterised by a flattening of the
sub-exponential behavior before it is cut-off exponentially fast. This
behaviour can be attributed to diffusion across streamlines into the
immobile regions in the wake of solid grains. 

In order to quantify these behaviors in a macroscale transport model,
we first develop a model for the statistics of the
Eulerian velocity magnitude in terms of the pore-size distribution. This model
relies on the assumption that the velocity profile in a single pore is
parabolic and that its maximum velocity is related to the
cross-sectional pore area through Poiseuille's law. Thus, we obtain an
estimate for the distribution of point-wise velocity magnitudes as
well as an estimate for the distribution of mean pore velocities. 
The distribution of velocity magnitudes is characterized by a flat
part at small velocities, which can be attributed to the flat
distribution characteristic for velocities in single pores, and an
exponential decay that can be attributed to the pore-size
distribution. The distribution of mean pore velocities is much
narrower than the distribution of the Eulerian velocity magnitude,
with smooth cut-offs at low and high velocity magnitudes. 

Then, we cast the advective-diffusive pore-scale particle motion in
terms of a time-domain random walk (TDRW), based on the fact that flow
velocities vary little over the characteristic pore length. We first
consider purely advective particle transitions such that the
distribution of residence times is given in terms of the distribution
of the Eulerian velocity magnitude. It scales as $t^{-3}$, which
implies a Levy-stable behavior for the particle travel time
distributions, and predicts a $Pe$--independent, persistent
$t^{-2}$--scaling of the breakthrough curves. This approach ignores 
completely the impact of diffusion on pore-scale particle
transport. Thus, in a next step, we account for the diffusive sampling of
flow velocities in single pores. Based on the assumption that pores
are in average much longer than wide, we approximate the effective
particle velocity over a pore by the average pore velocity. This
assumes that particles can sample the full pore velocity contrast
during a transition. Thus, the distribution of effective particle
velocities is set equal to the distribution of mean
pore-velocities. We also account for the diffusive cut-off of the
residence time distribution in case of diffusion-dominated
transitions. The resulting TDRW captures the intermediate tailing
of the breakthrough curves and the dependence on the P\'eclet number
despite these drastic simplifications. The flattening of the
breakthrough curve and asymptotic exponential cut-off is not accounted
for. As mentioned above, this feature can be attributed to purely
diffusive particle transitions across streamlines into the void space
in the wake of solid grains that connect adjacent pores. This
mechanism represents a trapping process, whose residence times
can be characterized by an exponential distribution which
decays on the diffusion time over a pore. The trapping rate can be
related to the volume fractions of mobile and immobile pore
regions and the inverse diffusion time. Trapping of particles is
quantified in the TDRW framework by a compound Poisson process
conditioned on the advective residence time. This approach captures
the full behavior of the breakthrough curves for all $Pe$. The
evolution of the particle density, or equivalently solute
distribution, is governed by an integro-partial differential
equation, whose memory kernels are related to the distribution of mean
pore velocities and the distribution of residence times in immobile
pores. Note that we have focused on global transport in the direction
of the mean flow. The presented approach can be generalised to account
for transverse mass transfer by determining the stochastic rules of
transverse particle motion, which here is quantified in an average sense through
tortuosity. 

In conclusion, we have identified and quantified the dominant
pore-scale advection and diffusion mechanism in a macroscale transport
model that captures pre-asymptotic non-Fickian transport. This
approach accounts for the impact of pore-scale heterogeneity on
transport on the REV scale, which manifests in tailing of the
breakthrough curves as an expression of pore-scale concentration
heterogeneity.   
Thus, while the concept of the REV is valid for the definition of macroscale
material properties such as porosity or permeability, it does not
imply that transport is in local physical equilibrium. 
The occurrence of advective breakthrough curve tailing in the intermediate regime depends
on the distance from the inlet and the distribution of characteristic advection time
scales, while the non-Fickian asymptotic tail due to trapping in
immobile zones depends on the characteristic diffusion time in the
immobile regions. We suggest that the identified transport mechanism
are of a general nature beyond the sand-like model medium under consideration here. More
complex porous media are characterized by broad distribution of pore
diameters and immobile porosity~\cite[][]{GMDC2008,bijeljic2013,
  gjetvaj2015}, which emphasise the
non-Fickian pre-asymptotic transport features and may delay the onset
of an asymptotic hydrodynamic dispersion regime. 

\begin{acknowledgments}
MD and JJH gratefully acknowledge the support of the European Research
Council (ERC) through the project MHetScale (617511) and the support
of the Spanish Ministry of Economy, Industry and Competitivity through
the project MECMAT (CGL2016-80022-R). MI gratefully acknowledges the financial support provided by AVL and
the computing facilities provided by the Warwick Centre for Scientific
Computing.
\end{acknowledgments}

\appendix
\section{Numerical simulations\label{app:of}}
The standard \textsf{OpenFOAM} solvers \textsf{simpleFoam} and
\textsf{scalarTransportFoam} have been used to perform, respectively,
flow and transport simulations. The latter has been modified to allow
for runtime co-processing of volumetric and PDF
information. 


A blended forward Euler/Crank-Nicolson discretisation in time has been
used that ensures $\Delta t < {\Delta x^{2}} / {D}$ for accurate time
resolution. A second-order upwind scheme is used for
advection to minimize numerical dispersion, a second-order least
square approach is used for computing gradients and a central scheme
for diffusive fluxes.

The simulations are performed on an irregular grid to represent well the
grain boundaries. Details on the mesh-generation and numerical method are given in
\cite{icardi2014pore}. In Sections III and IV of this paper, the
interested reader finds a detailed grid independence study of the flow
and transport simulations. The number of mesh cells used for the
numerical solution is $41 \cdot 10^6$. The largest cell size is
$\ell_c = 1/200$, the smallest $\ell_c 1/1600$ of the domain size. Simulations were performed on
128 Intel Xeon E5-2630 2.4GHz cores and, due to the significant tailing, have to be run for long
times, resulting in approximately 48h of simulation time per each
run. 

Regarding the issue of numerical dispersion, for the numerical
simulations used in the manuscript, the maximum grid
P\'eclet number $Pe_g = \ell_c v /D$ varies between $Pe_g \approx 5$
for the $Pe = 10^3$ case and $Pe_g \approx 0.15$ for the $Pe = 30$
case. Note that these are the maximum grid Peclet numbers, which are
obtained in the pore centers, where the resolution is coarsest and
velocities are relatively high, of the order of the mean velocity. As both the mesh-size and velocity
vary, the grid Peclet numbers close to grains and in the wake of the
grains, where the resolutions are high (i.e., small $\ell_c$) and velocities
are low, the grid Peclet numbers are $Pe_g \ll 1$. Also, note that for
the breakthrough curves under consideration
(e.g., Figure~\ref{fig:btc}), the effect of hydrodynamic dispersion due
to flow variability (in high flow velocities), which is of the order of $\overline v \ell_0$,
overshadows the possible effect of numerical dispersion in the pore
centers, which would otherwise only be visible in the initial decay of the breakthrough curve. The tailings
in the two intermediate regimes are caused by variability in low
velocities (high velocities give rise to the hydrodynamic dispersion
effect) and dominant diffusion in immobile zones, for which the local
grid P\'eclet numbers are much smaller than 1. For this reason, we are
confident that the numerical simulations accurately reflect the actual
physical mass transfer processes in the porous medium. 
\section{Kernels\label{app:kernels}}
We derive here expressions~\eqref{kernels} and~\eqref{kernelsmrmt}. 
To this end, we consider the $i$th moments of the Laplace transform
of the $\psi(x,t)$
\begin{align}
\mu^\ast_i(\lambda) = \int dx x^i \psi^\ast(x,\lambda). 
\end{align}
The Laplace transforms of the kernels~\eqref{kdef} are then given in terms of the $\mu_i^\ast(\lambda)$ as
\begin{subequations}
\label{app:kdef}
\begin{align}
\nu^\ast(\lambda) &= \frac{\mu_1^\ast(\lambda)}{1 - \psi^\ast(\lambda)}
\\
\kappa^\ast(\lambda) &= \frac{1}{2} \frac{\mu_2^\ast(\lambda)}{1 - \psi^\ast(\lambda)}.  
\end{align}
\end{subequations}

We obtain by using the Laplace transform of~\eqref{psim}
\begin{align}
\mu_{1,m}^\ast(\lambda) &= \frac{\ell}{\chi}\int\limits_0^\infty dv \frac{v^2 \tau_v p_m(v)}{\langle v_e \rangle} \frac{\lambda \psi_0^\ast(\lambda|v)}{1-\psi_0^\ast(\lambda|v)} \left[1-\psi_0^\ast(\lambda|v) \right]  
\\
\mu_{2,m}^\ast(\lambda) &= \int\limits_0^\infty dv \frac{2 D + v \ell}{\chi^2} \frac{v \tau_v p_m(v)}{\langle v_e \rangle} \frac{\lambda \psi_0^\ast(\lambda|v)}{1-\psi_0^\ast(\lambda|v)} \left[1-\psi_0^\ast(\lambda|v) \right]  
\end{align}
Inserting the Laplace transform of the exponential $\psi_0(t|v)$ then gives 
\begin{align}
\mu_{1,m}^\ast(\lambda) &= \frac{\ell}{\chi}\int\limits_0^\infty dv \frac{v^2 p_m(v)}{\langle v_e \rangle} 
\left[1-\psi_0^\ast(\lambda|v) \right]  
\label{app:mu1}
\\
\mu_{2,m}^\ast(\lambda) &= \int\limits_0^\infty dv \frac{2 D + v \ell}{\chi^2} \frac{v p_m(v)}{\langle v_e \rangle} \left[1-\psi_0^\ast(\lambda|v) \right].
\label{app:mu2}  
\end{align}
Inserting these expressions into~\eqref{app:kdef} gives~\eqref{kernels}. 

Similarly, we obtain by using~\eqref{psimrmt} in~\eqref{app:kdef}
\begin{align}
\mu_{1}^\ast(\lambda) &= \frac{\ell}{\chi}\int\limits_0^\infty dv \frac{v^2 \tau_v p_m(v)}{\langle v_e \rangle} \frac{\lambda \Phi^\ast(\lambda) \psi_0^\ast[\lambda \Phi^\ast(\lambda)|v]}{1-\psi_0^\ast[\lambda\Phi^\ast(\lambda)|v]} \frac{1-\psi_0^\ast[\lambda \Phi^\ast(\lambda)|v]}{\Phi^\ast(\lambda)}  
\\
\mu_{2}^\ast(\lambda) &= \int\limits_0^\infty dv \frac{2 D + v \ell}{\chi^2} \frac{v \tau_v p_m(v)}{\langle v_e \rangle} \frac{\lambda \Phi^\ast(\lambda) \psi_0^\ast[\lambda \Phi^\ast(\lambda)|v]}{1-\psi_0^\ast[\lambda\Phi^\ast(\lambda)|v]} \frac{1-\psi_0^\ast[\lambda \Phi^\ast(\lambda)|v]}{\Phi^\ast(\lambda)},  
\end{align}
where we defined $\Phi^\ast(\lambda) = 1 + \lambda^{-1} \gamma[1 - p_f^\ast(\lambda)]$. 
Inserting the Laplace transform of the exponential $\psi_0(t|v)$ then gives 
\begin{align}
\mu_1^\ast(\lambda) &= \frac{\ell}{\chi}\int\limits_0^\infty dv \frac{v^2 p_m(v)}{\langle v_e \rangle} 
\frac{1-\psi_0^\ast[\lambda \Phi^\ast(\lambda)|v]}{\Phi^\ast(\lambda)}  
\\
\mu_2^\ast(\lambda) &= \int\limits_0^\infty dv \frac{2 D + v \ell}{\chi^2} \frac{v p_m(v)}{\langle v_e \rangle} \frac{1-\psi_0^\ast[\lambda \Phi^\ast(\lambda)|v]}{\Phi^\ast(\lambda)}.   
\end{align}
By comparison with~\eqref{app:mu1} and~\eqref{app:mu2}, we observe that 
\begin{align}
\mu_1^\ast(\lambda) &= \frac{\mu_{1,m}^\ast[\lambda\Phi^\ast(\lambda)]}{\Phi^\ast(\lambda)}  
\\
\mu_2^\ast(\lambda) &= \frac{\mu_{2,m}^\ast[\lambda\Phi^\ast(\lambda)]}{\Phi^\ast(\lambda)}.   
\end{align}
Inserting these expressions into~\eqref{app:kdef} gives~\eqref{kernelsmrmt}.
\section{Transition time distribution\label{app:psi}}
Here we derive expression~\eqref{psimrmt} for the Laplace transform of the transition time PDF for the case of 
heterogeneous advection and trapping. The trapping time, the second term in~\eqref{taumrmt} is a compound Poisson process, whose distribution is $\psi_f(t|\tau_m)$, see~\eqref{psicp}.  
Thus, for a given $\tau_m$, the Laplace transform of the joint PDF of transition length and time is 
\begin{align}
\label{app:psiast}
\psi^\ast(\lambda|\tau_m) = \exp\left(- \left\{\lambda + \gamma\left[1 - p_f^\ast(\lambda) \right] \right\} \tau_m - \lambda \tau_m \right). 
\end{align}
For a given $v$, then the distribution of transition times is obtained by averaging of~\eqref{app:psiast},
\begin{align}
\psi^\ast(\lambda|v) &= \int\limits_0^\infty dt \exp\left(- \left\{\lambda + \gamma\left[1 - p_f^\ast(\lambda) \right] \right\} t - \lambda t \right) \psi_0(t|v) 
\nonumber\\
&= \psi_0^\ast(\lambda + \gamma [1 - p_f^\ast(\lambda)]). 
\end{align}
The joint PDF of transition length and time is thus
\begin{align}
\psi^\ast(\lambda) = \int\limits_0^\infty dv \frac{v p_m(v)}{\langle v_e \rangle} \psi_0^\ast(\lambda + \gamma [1 - p_f^\ast(\lambda)]) =  \psi_m^\ast(\lambda + \gamma [1 - p_f^\ast(\lambda)]). 
\end{align}
%
\section{Breakthrough curves\label{app:fptd}}
We determine here the solution for the breakthrough curve corresponding to the governing equation~\eqref{mrmt}.
As at the beginning of Section~\ref{sec:CTRW} for the ADE-model, we consider a semi-infinite domain, which is a 
good approximation for $Pe \gg 1$, which is the case for all the scenarios under consideration here. Furthermore, 
we notice that Laplace transform of~\eqref{mrmt} is given by 
\begin{align}
\label{app:mrmt}
\lambda c_m^\ast(x,\lambda)  + \langle u_1 \rangle \frac{\partial
  c_m^\ast(x,\lambda)}{\partial x} - \mathcal D \frac{\partial^2 c_m^\ast(x,\lambda)}{\partial
  x^2} = 0,
\end{align}
where we used~\eqref{cim} and the fact that the initial concentration is $0$. Furthermore, we defined 
\begin{align}
\varphi^\ast(\lambda) = \lambda^{-1} \left[1 - p_f^\ast(\lambda) \right]. 
\end{align}
Note that~\eqref{app:mrmt} is identical in form to the Laplace transform of~\eqref{ademac}. Thus, the solution can be expressed in terms of the 
Laplace transform of~\eqref{eq:ade_analytic} as~\cite[][]{DCSB2004} 
\begin{align}
\label{app:ade_analytic:l}
f^\ast(\lambda,x) = \frac{1}{\lambda} - \frac{1}{\lambda} \exp\left[- \frac{x \langle u_1 \rangle}{2 D} \left(\sqrt{1 + 4 \frac{\lambda D [1+\gamma \varphi^\ast(\lambda)]}{\langle u_1 \rangle^2}} \right) -1 \right]. 
\end{align}
The solutions of~\eqref{mrmt} for the breakthrough curves displayed in Figure~\ref{fig:btc4} are obtained by numerical inverse Laplace transform 
of~\eqref{app:ade_analytic:l}. In the limit $\lambda D/\langle u_1 \rangle^2 \ll 1$, we can expand the latter as
\begin{align}
f^\ast(\lambda,x) = \frac{x [1+\gamma \varphi^\ast(\lambda)]}{\langle u_1 \rangle} + \dots, 
\end{align}
where the dots denote subleading contributions. Thus, we obtain for $f(t,L) \equiv f(t)$ the long-time approximation~\eqref{f:approx}.  
%
\section{Trapping time distribution\label{app:trap}}
In order to make an estimate for the distribution of trapping time, we approximate diffusion into immobile pores
as a $1$--dimensional first passage problem. Particles are injected at $z = 0$ and diffuse to the outlet at $z = \ell$. 
 The particle distribution $g(z,t)$ then follows the diffusion equation
\begin{align}
\label{app:diffeq}
\frac{\partial g(z,t)}{\partial t} - D \frac{\partial^2
  g(z,t)}{\partial z^2} = 0
\end{align}
for the boundary conditions
\begin{align}
\label{app:bc}
- D \frac{\partial g(z = 0,t)}{\partial z} = \delta(t), && g(z = \ell,t) = 0, 
\end{align}
and the initial condition $g(z,t = 0) = 0$. The distribution of first passage times at $z = \ell$ 
 is given by the flux over the boundary as 
\begin{align}
\label{app:f}
p_f(t) =  - D \frac{g(z = \ell,t)}{\partial z}. 
\end{align}
The solution to~\eqref{app:diffeq} reads in Laplace space as
\begin{align}
g^\ast(z,\lambda) = \frac{\sinh\left[\sqrt{\frac{\lambda}{D}} (\ell - z) \right]}{\sqrt{\lambda D} \cosh\left(\sqrt{2 \lambda \tau_D} \right)}. 
\end{align}
Using the latter in the Laplace transform of~\eqref{app:f} gives
\begin{align}
\label{f}
p_f^\ast(\lambda) = \frac{1}{\cosh\left(\sqrt{2 \lambda \tau_D}\right)}.
\end{align}
For $\lambda \tau_D \ll 1$~\eqref{f} can be approximated by 
\begin{align}
p_f^\ast(\lambda) \approx \frac{1}{1 + \lambda \tau_D},
\end{align}
whose inverse Laplace transform is given by~\eqref{psif}. 
\section{Governing equation under heterogeneous advection and trapping\label{app:advmrmt}}
In order to derive~\eqref{cadvmrmt3}, we note that the governing equation for the Laplace transform $\overline c^\ast(x,\lambda)$ 
follows from~\eqref{nlade} and~\eqref{kernelsmrmt} as
\begin{align}
\label{cadvmrmt}
\lambda \overline c^\ast(x,\lambda) + \left(\nu^\ast_m[\lambda \Phi^\ast(\lambda)] \frac{\partial}{\partial x} - 
\kappa^\ast_m[\lambda \Phi^\ast(\lambda)] \frac{\partial^2}{\partial x^2} \right) \frac{\overline c^\ast(x,\lambda)}{\Phi^\ast(\lambda)} = c(x,t=0). 
\end{align}
The Laplace transform of the immobile concentration~\eqref{cim} is given by 
\begin{align}
c_{im}^\ast(x,\lambda) = \gamma \lambda^{-1}\left[1 - p_f^\ast(\lambda)\right] c_m^\ast(x,\lambda), 
\end{align}
such that the total concentration is $\overline c^\ast(x,\lambda) = \Phi^\ast(\lambda) c_m^\ast(x,\lambda)$. Thus,~\eqref{cadvmrmt}
can be written in terms of the mobile concentration $c^\ast_m(x,\lambda)$ as
\begin{align}
\label{cadvmrmt2}
\lambda \Phi^\ast(\lambda) c_m^\ast(x,\lambda)  + \left(\nu^\ast_m[\lambda \Phi^\ast(\lambda)] \frac{\partial}{\partial x} - 
\kappa^\ast_m[\lambda \Phi^\ast(\lambda)] \frac{\partial^2 }{\partial x^2} \right) c_m^\ast(x,\lambda)  = c(x,t=0). 
\end{align}
Its inverse Laplace transform gives~\eqref{cadvmrmt3}. 
\bibliographystyle{jfm}

\end{document}